\documentclass[journal]{IEEEtran}

\normalsize

\ifCLASSINFOpdf
\else
\fi

\usepackage[font=footnotesize]{subfig}
\usepackage{array}
\usepackage{fixltx2e}
\usepackage[cmex10]{amsmath}
\usepackage[font=footnotesize, margin=0.0cm]{caption}
\usepackage{graphicx}
\usepackage{amsmath}
\usepackage{amsthm, amssymb}
\usepackage[overload]{empheq}
\usepackage{commath}
\usepackage{caption}
\usepackage{multirow}
\usepackage{algorithm}
\usepackage{algorithmic}
\newcommand{\RN}[1]{%
  \textup{\uppercase\expandafter{\romannumeral#1}}%
  }
\usepackage{mathtools}

\DeclareMathAlphabet{\mathpzc}{OT1}{pzc}{m}{it}
\usepackage{cite}
\usepackage{color}
\usepackage[utf8]{inputenc}
\usepackage[english]{babel}

\newtheorem{remark}{Remark}

\newtheorem{proposition}{Proposition}
\usepackage{url}
\usepackage{setspace}
\usepackage{changebar}

\graphicspath{{../fig/}}   

\begin{document}
%
\title{Latency Minimization for Intelligent Reflecting Surface Aided Mobile Edge Computing}
%
%

\author{Tong Bai, \IEEEmembership{Member,~IEEE},
		Cunhua Pan, \IEEEmembership{Member,~IEEE},
		Yansha Deng, \IEEEmembership{Member,~IEEE}, \\
		Maged Elkashlan, \IEEEmembership{Member,~IEEE},
		Arumugam Nallanathan, \IEEEmembership{Fellow,~IEEE}
		and Lajos Hanzo, \IEEEmembership{Fellow,~IEEE}
\thanks{T. Bai, C. Pan, M. Elkashlan, and A. Nallanathan are with the School of Electronic Engineering and Computer Science, Queen Mary University of London, London, E1 4NS, U.K. (e-mail: t.bai@qmul.ac.uk, c.pan@qmul.ac.uk, maged.elkashlan@qmul.ac.uk, a.nallanathan@qmul.ac.uk).}
\thanks{Y. Deng is with the Department of Engineering, King’s College London, London, WC2R 2LS, U.K. (e-mail: yansha.deng@kcl.ac.uk).}
\thanks{L. Hanzo is with the School of Electronics and Computer Science, University of Southampton, Southampton, SO17 1BJ, U.K. (e-mail: lh@ecs.soton.ac.uk).}
\thanks{(Corresponding author: Cunhua Pan and Tong Bai)}
}

\maketitle
\IEEEpeerreviewmaketitle

\vspace{-1.0cm}
\begin{abstract}
Computation off-loading in mobile edge computing (MEC) systems constitutes an efficient paradigm of supporting resource-intensive applications on mobile devices. However, the benefit of MEC cannot be fully exploited, when the communications link used for off-loading computational tasks is hostile.
Fortunately, the propagation-induced impairments may be mitigated by intelligent reflecting surfaces (IRS), which are capable of enhancing both the spectral- and energy-efficiency.
Specifically, an IRS comprises an IRS controller and a large number of passive reflecting elements, each of which may impose a phase shift on the incident signal, thus collaboratively improving the propagation environment.
In this paper, the beneficial role of IRSs is investigated in MEC systems, where single-antenna devices may opt for off-loading a fraction of their computational tasks to the edge computing node via a multi-antenna access point with the aid of an IRS.
Pertinent latency-minimization problems are formulated for both single-device and multi-device scenarios, subject to practical constraints imposed on both the edge computing capability and the IRS phase shift design.
To solve this problem, the block coordinate descent (BCD) technique is invoked to decouple the original problem into two subproblems, and then the computing and communications settings are alternatively optimized using low-complexity iterative algorithms. It is demonstrated that our IRS-aided MEC system is capable of significantly outperforming the conventional MEC system operating without IRSs. Quantitatively, about $20~\%$ computational latency reduction is achieved over the conventional MEC system in a single cell of a $300~\rm{m}$ radius and $5$ active devices, relying on a $5$-antenna access point.
\end{abstract}

\begin{IEEEkeywords}
Intelligent reflecting surface, mobile edge computing, latency minimization.
\end{IEEEkeywords}

\section{Introduction}
\label{sec:Introduction}

\subsection{Motivation and Scope}
In the Internet-of-Things (IoT) era, myriads of machines and sensors are envisioned to be connected \cite{palattella2016internet}. However, since these devices typically have limited computing capabilities, resource-intensive applications cannot be readily supported by these devices due to their resultant excessive computational latency. Aimed at tackling this issue, powerful computing nodes can be deployed at the edge of the network (typically co-located with the access points (APs)) \cite{tropic}. As a benefit, the computational latency of these resource-intensive applications can be reduced, by employing both local computing on the devices and edge computing for processing these computational tasks, provided that these tasks can be successfully off-loaded. This paradigm is referred to as mobile edge computing (MEC)  \cite{barbarossa2014communicating,shi2016edge,mao2017survey,
zhang2013energy,wang2016mobile,ren2018latency,bai2019energy,sardellitti2015joint,
chen2015efficient,lyu2017optimal,dai2018joint,ouyang2018follow}.
At the time of writing, the potential of this MEC paradigm has not been fully exploited, predominantly because the computation off-loading link is far from perfect. For example, the devices located at the cell edge typically suffer from a low off-loading success rate, and/or their computation off-loading may impose higher latency than computing their tasks locally. Hence these devices have to rely on their own computing resources, which is however often incapable of supporting resource-intensive applications. Therefore, it is imperative to improve the performance of MEC systems from a communications perspective.

The recent advances in programmable meta-materials \cite{cui2014coding} facilitate the construction of intelligent reflecting surfaces (IRSs) \cite{di2019smart} for enhancing both the spectral- and energy-efficiency of wireless communications. Specifically, an IRS is comprised of an IRS controller and a large number of passive reflecting elements. Under the instructions of the IRS controller, each IRS reflecting element is capable of adjusting both the amplitude and the phase of the signals reflected, thus collaboratively modifying the signal propagation environment. The gain attained by IRSs is based on the combination of both the virtual array gain and the reflection-aided beamforming gain. To elaborate, the virtual array gain can be achieved by combining both the direct and IRS-reflected signals, while the reflection-aided beamforming gain is realized by proactively controlling the phase shift induced by the IRS elements. By beneficially combining these two types of gains, the IRS becomes capable of boosting the devices' off-loading success rate, hence improving the potential of MEC systems.
In this treatise, our attention is focused on investigating the role of IRSs in MEC systems.

\subsection{Related Works}


\subsubsection{Design of Mobile Edge Computing Systems}

At the current state-of-the-art, MEC systems can be categorized into \cite{mao2017survey}: single-user \cite{zhang2013energy,wang2016mobile,ren2018latency,bai2019energy} and multi-user systems \cite{sardellitti2015joint,chen2015efficient,lyu2017optimal,dai2018joint,ouyang2018follow}. Among the design metrics of single-user MEC systems, the computation off-loading strategy plays a crucial role. More explicitly, the binary off-loading strategy of \cite{zhang2013energy} was proposed to decide whether the task is executed locally at the mobile device or remotely at the edge-cloud node. By contrast, Wang \emph{et al.} \cite{wang2016mobile} conceived a partial off-loading scheme for data-partitioning oriented applications, where a fraction of the data can be processed at the mobile device, while the rest at the edge. However, in realistic multi-user systems, inter-user interference is imposed both on the radio communications link and on the computing node at the edge, which may erode the overall performance of the MEC system. In order to cope with this hindrance, Sardellitti \emph{et al.} \cite{sardellitti2015joint} jointly optimized the transmit precoding matrices and the computational resources allocated to each user in a multi-cell multi-user scenario, while Sheng \emph{et al.} \cite{sheng2019energy} proposed an energy-efficient algorithm to optimize the resource allocation of terminals, radio access networks, and edge servers in a multi-carrier scenario. For the system where the devices have to make their off-loading decisions locally, Chen \emph{et al.} \cite{chen2015efficient} provided a distributed joint computation off-loading and channel selection policy relying on classic game theory. Recently, a specific user association scheme was also developed for multi-user systems served by multiple edge computing nodes \cite{dai2018joint}, while a mobility-aware dynamic service scheduling algorithm was proposed for MEC systems \cite{ouyang2018follow}.
Furthermore, Yang \emph{et al.}  \cite{yang2020federated} conceived an over-the-air computation aided federated learning algorithm for reducing both the latency and the power consumption, and for preserving the users' privacy in MEC systems.
At the time of writing, the computation offloading issue of the devices in the face of hostile communications environments has not been well addressed. Against this background, in this paper the performance is improved by invoking IRSs. Let us now continue by reviewing the relevant research contributions on IRSs as follows.

\subsubsection{Intelligent Reflecting Surface Aided Wireless Networks}
In order to explore the benefits of IRSs in wireless communications, extensive research efforts have been invested into their ergodic capacity analysis \cite{han2019large}, channel estimation \cite{zheng2019intelligent}, and practical reflection phase shift modeling \cite{abeywickrama2019intelligent}, as well as into the associated phase shift design \cite{wu2018intelligent,wu2019beamforming,
guo2019weighted,ye2019joint,pan2019intelligent,zhou2019intelligent,huang2019indoor,yang2019intelligent}. Specifically, a joint design of the IRS phase shift and of the precoding at the AP was proposed for minimizing the transmit power, while maintaining the target receive signal-to-interference-plus-noise ratio (SINR) \cite{wu2018intelligent}, relying on the sophisticated techniques of the semidefinite relaxation and of alternating optimization. These investigations were then extended to the more practical discrete phase shift setting \cite{wu2019beamforming}.
However, the excessive computational complexity of the algorithm developed in \cite{wu2018intelligent} prohibits its application in large-scale IRSs. In order to reduce the complexity, Guo \emph{et al.} \cite{guo2019weighted} proposed three low-complexity algorithms, while Pan \emph{et al.} \cite{pan2019intelligent} provided a pair of majorization-minimization (MM) algorithms and complex circle manifold methods for multi-cell scenarios.
Furthermore, in order to reduce the overhead during the IRS channel estimation, Yang \emph{et al.} \cite{yang2019intelligent} grouped the IRS elements, where each group shares the same phase shift coefficient, and optimized the power allocation and phase shift in orthogonal frequency division multiplexing (OFDM)-based wireless systems.
Apart from the conventional communications scenarios, the role of IRSs was also investigated both in terms of improving physical-layer security \cite{8723525,8743496,xu2019resource,chen2019intelligent}, and simultaneous wireless information and power transfer (SWIPT) \cite{wu2019weighted,pan2019swipt}, where substantial gains were achieved. These impressive research contributions inspired us to exploit the beneficial role of IRSs in MEC systems.

\subsection{Contributions and Organizations}

Our main contributions are the employment of IRSs in MEC systems, and the joint design of computing and communications for minimizing the computational latency of IRS-aided MEC systems, detailed as follows.
\begin{itemize}
\item \emph{New IRS-aided MEC system design and latency minimization problem formulation:} In order to further exploit the potential of MEC systems, we first propose IRS-aided MEC systems, for assisting the computational task off-loading of mobile devices. A latency-minimization problem is formulated for multi-device scenarios, which optimizes the computation off-loading volume, the edge computing resource allocation, the multi-user detection (MUD) matrix, and the IRS phase shift, subject to both the total edge computing capability and to the IRS phase shift constraints. Owing to the coupling effect of multiple optimization variables, the latency-minimization problem cannot be solved directly. Hence, relying on the block coordinate descent (BCD) technique, the original problem is decoupled into two subproblems for alternatively optimizing computing and communications settings.
\item \emph{Computing design:}
Given a fixed communications setting, we decouple the computation off-loading volume and the edge computing resource allocation, again using the BCD technique. Our analysis reveals that given a fixed edge computing resource allocation, the optimal computation off-loading volume can be determined by assuming the equivalence of the latency induced by local computing and by edge computing. Given a fixed computation off-loading volume, the subproblem is proved to be a convex problem, and the optimal edge computing resource can be found by relying on the KKT conditions and on the classic bisection search method.
\item \emph{Communications design:}
Given a computing setting, the objective function (OF) becomes available in a non-convex sum-of-ratios form, which cannot be solved using the algorithms developed in \cite{wu2018intelligent,guo2019weighted,pan2019intelligent,pan2019swipt}.
To tackle this challenge, this problem is transformed to an equivalent parameterized form by introducing auxiliary variables. Our analysis reveals that this equivalent form can be decomposed into a series of tractable subproblems. Then, an iterative algorithm is developed to find the solution. In each iteration, the auxiliary variables are updated using the modified Newton's method, while upon reformulating this series of tractable subproblems by exploiting the equivalence between the weighted sum-rate maximization problem and the weighted mean square error (MSE) minimization problem, closed-form expressions are provided for the MUD matrix and for the IRS phase shift, using the weighted minimum MSE method and MM algorithm, respectively. Our analysis reveals that the proposed algorithm exhibits a low complexity.
\item \emph{Study of the single-device scenario:} In order to complete the investigations, the single-device scenario is also studied, where neither the edge computing resource allocation nor the multi-user interference has to be considered. A low-complexity iterative algorithm is proposed by simplifying the algorithm developed in the aforementioned multi-device scenario.
\item \emph{Numerical validations and evaluations:}  The numerical results verify the convergence of the proposed algorithms, and quantify the performance of our IRS-aided MEC system in terms of its latency in diverse simulation environments.
\end{itemize}

The rest of the paper is organized as follows. In Section~\ref{sec:System Model}, we establish the system model and formulate the latency minimization problem. The solution of this latency minimization problem is provided in Section~\ref{sec:Solution to Problem}. In Section~\ref{sec:Special Case Study}, we investigate the solution of the special case, where a single device is served by the MEC system. Our numerical results are discussed in Section~\ref{sec:Numerical Results}. Finally, our conclusions are offered in Section~\ref{sec:Conclusions}.

\emph{Notation:} In this paper, scalars are denoted by italic letters. Boldface lower- and upper-case letters denote vectors and matrices, respectively; $\mathbb{C}^{M\times N}$ represents the space of $M\times N$ complex matrices; $\pmb{I}_N$ denotes an $N\times N$ identity matrix; $j$ denotes the imaginary unit, i.e. $j^2 = -1$.
The maths operations used throughout the paper are summarized in Table~\ref{tbl:math_operations}.
\begin{table}[h!]\small
\begin{center}
\caption{Math operations}
\label{tbl:math_operations}
\begin{tabular}{  l | r }
\hline
Notation & Operation \\ \hline \hline
${\pmb{x}}^T$ &  the transpose of $\pmb{x}$ \\ \hline
${\pmb{X}}^T$ &  the transpose of $\pmb{X}$ \\ \hline
${\pmb{x}}^*$ &  the complex conjugate of $\pmb{x}$ \\ \hline
${\pmb{X}}^*$ &  the complex conjugate of $\pmb{X}$ \\ \hline
${\pmb{x}}^H$ &  the Hermitian transpose of $\pmb{x}$ \\ \hline
${\pmb{X}}^H$ &  the Hermitian transpose of $\pmb{X}$ \\ \hline
${\pmb{X}}^{-1}$ & the inverse of $\pmb{X}$ \\ \hline
$\bigodot$ & the Hadamard product \\ \hline
$\Re\{\cdot\}$ & the real part of a complex number \\ \hline
$\text{arg}\{ \cdot \}$ & the argument of a complex number \\ \hline
$| \cdot |$ & the absolute value of a scalar \\ \hline
$\| \cdot \|$ & the 2-norm of a vector \\ \hline
$\lfloor \cdot \rfloor$ & the floor of a scalar \\ \hline
$\lceil \cdot \rceil$ & the ceiling of a scalar \\ \hline
\multirow{2}{*}{$\text{diag}(\pmb{x})$}
&  the diagonal matrix where \\
& the diagonal elements are $\pmb{x}$ \\ \hline
\multirow{2}{*}{$\text{diag}(\pmb{X})$}
& the vector whose elements are \\
& the diagonal elements of $\pmb{X}$ \\ \hline
\multirow{2}{*}{$ \mathcal{CN}(0,\sigma^2)$}
& Circularly Symmetric Complex Gaussian \\
& associated with zero-mean and variance $\sigma^2$ \\ \hline
\end{tabular}
\end{center}
\end{table}

\section{System Model and Problem Formulation}
\label{sec:System Model}

In this section, our system model is elaborated on, from both communications and computing perspectives. Following this, a latency-minimization problem is formulated for our IRS-aided MEC system, detailed as follows.

\subsection{Communications Model}
\begin{figure}[t!]\center
\includegraphics[width= 0.5 \textwidth]{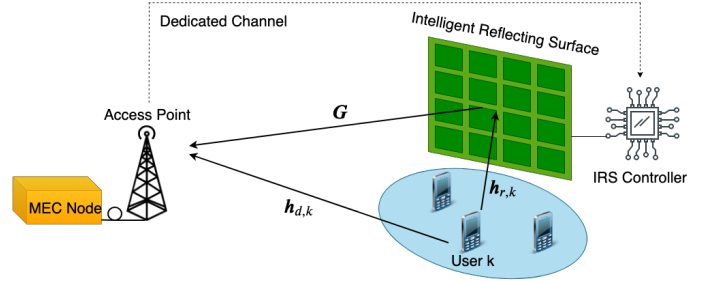}
\caption{Illustration of the system model, where a $N$-element intelligent reflecting surface (IRS) assists the computation off-loading of $K$ single-antenna devices to the edge computing node via the $M$-antenna access point.}
\label{fig:system_model}
\end{figure}

As shown in Fig.~\ref{fig:system_model}, we consider an MEC system operating in a single-cell scenario, where $K$ single-antenna devices may opt for off-loading a certain fraction of or all of their computational tasks to an edge computing node via an $M$-antenna AP through the wireless transmission link. The edge computing node and  the AP are assumed to be co-located and connected using high-throughput low-latency optical fiber. Then, the latency imposed by the data communication between the AP and the edge computing node is deemed to be negligible. An IRS comprised of $N$ reflecting elements is placed in the cell for assisting the devices' computation off-loading. We assume that both the antenna spacing at the AP and the element spacing of the IRS are high enough so that the small-scale fading associated both with two different antennas and with two different reflecting elements is independent, respectively.

The equivalent baseband channels spanning from the $k$-th device to the AP, and from the $k$-th device to the IRS, as well as from the IRS to the AP are denoted by $\pmb{h}_{d,k} \in \mathbb{C}^{M\times 1}$, $\pmb{h}_{r,k} \in \mathbb{C}^{N \times 1}$, and $\pmb{G} \in \mathbb{C}^{M \times N}$, respectively.
These channels are assumed to be perfectly estimated\footnote{Naturally, this assumption is idealistic. Hence the algorithm developed in this paper can be deemed to represent the best-case bound for the latency performance of realistic scenarios.} and quasi-static, hence remaining near-constant when devices are scheduled for off-loading their computational tasks.
As for the IRS, we simply set the amplitude reflection coefficient to $1$ for all reflection elements and denote the phase shift coefficient vector by
 $\pmb{\theta} = [\theta_1, \theta_2, \ldots, \theta_N]^T$, where $\theta_n \in [0,2\pi)$ for all $n\in \{1,2,\ldots,N\}$.\footnote{Due to the associated hardware limitations, only a limited number of discrete phase shifts can be provided for each IRS element in practice \cite{wu2019beamforming}. Our proposed algorithms provide the best-case bound for the latency of realistic scenarios. The phase-quantization effects are evaluated in Section~\RN{5}-A3.} Then, we have the reflection-coefficient matrix of the IRS $\pmb{\Theta} = \text{diag}\big\{ e^{j\theta_1}, e^{j\theta_2}, \ldots, e^{j\theta_N} \big\}$, where $j$ represents the imaginary unit.
It is assumed that the IRS phase shift setting is calculated at the AP in accordance with both the channel and computing dynamics, which is then sent to the IRS controller along the dedicated channel.
The composite device-IRS-AP channel is modeled as a concatenation of the device-IRS link, the IRS reflection characterized by its phase shift, and the IRS-AP link.

Here, it is assumed that the computation off-loading of the $K$ devices takes place over a given frequency band $B$ within the same time resource.
Upon denoting the off-loading power, and off-loading signal of the $K$ devices, as well as the noise vector by $p_t$, $\pmb{s} = [s_1,s_2,\ldots,s_K]^T$, and $\pmb{n} = [n_1,n_2, \ldots, n_M]^T$, respectively,  the signal $\pmb{y} \in \mathbb{C}^{M\times 1}$ received at the AP is readily formulated as
\begin{IEEEeqnarray}{rCl}
\pmb{y}
 =  \sqrt{p_t} \pmb{H} \pmb{s} + \pmb{n}  =  \sqrt{p_t} \sum^{K}_{k=1} \big( \pmb{h}_{d,k} + \pmb{G} \pmb{\Theta} \pmb{h}_{r,k} \big) s_k + \pmb{n},
\end{IEEEeqnarray}
where we assume $n_m \sim \mathcal{CN}(0,\sigma^2)$ for $m = 1,2,\ldots,M$. Furthermore, we define $\pmb{h}_k \triangleq \pmb{h}_{d,k} + \pmb{G} \pmb{\Theta} \pmb{h}_{r,k}$ and $\pmb{H} \triangleq \big[\pmb{h}_1, \pmb{h}_2, \ldots, \pmb{h}_K \big]$.
As a computational complexity compromise at the AP, a linear MUD technique is invoked. Upon denoting the MUD matrix by $\pmb{W} \in \mathbb{C}^{M \times K}$, the signal recovered at the AP is obtained as
\begin{IEEEeqnarray}{rCl}
\hat{\pmb{s}} = \pmb{W}^H \pmb{y} = \pmb{W}^H (\sqrt{p_t} \pmb{H} \pmb{s} + \pmb{n}).
\end{IEEEeqnarray}
As for the $k$-th device, its recovered signal is formulated as
\begin{IEEEeqnarray}{rCl}\label{eqn:recovered_signal}
\hat{s}_k = \pmb{w}_k^H \Bigg[ \sqrt{p_t} \sum^{K}_{j=1} \big( \pmb{h}_{d,j} + \pmb{G} \pmb{\Theta} \pmb{h}_{r,j} \big) s_j + \pmb{n} \Bigg],
\end{IEEEeqnarray}
where $\pmb{w}_k$ is the $k$-th column of the matrix $\pmb{W}$.
Then, the SINR of the $k$-th device's signal recovered is given by
\begin{equation}\label{eqn:SINR}
\gamma_k(\pmb{w}_k,\pmb{\theta}) = \frac{p_t \big|\pmb{w}_k^H \big( \pmb{h}_{d,k} + \pmb{G} \pmb{\Theta} \pmb{h}_{r,k} \big) \big|^2}{p_t \sum^K_{j=1, j\neq k} \big|\pmb{w}_k^H \big( \pmb{h}_{d,j} + \pmb{G} \pmb{\Theta} \pmb{h}_{r,j} \big)\big|^2 + \sigma^2 |\pmb{w}_k^H|^2 }.
\end{equation}
Accordingly, upon assuming a perfect capacity-achieving transmission scheme is invoked, we arrive at the maximum achievable computation off-loading rate of the $k$-th device, formulated as
\begin{equation}\label{eqt:throughput}
R_k(\pmb{w}_k,\pmb{\theta}) = B \log_2 \big[ 1 + \gamma_k(\pmb{w}_k,\pmb{\theta}) \big].
\end{equation}

\subsection{Computing Model}

We consider the data-partitioning based application of \cite{wang2016mobile}, where a fraction of the data can be processed locally, while the other part can be off-loaded to the edge node. The computing model is detailed for the local and edge computing as follows.
\begin{itemize}
\item \emph{Local computing:} For the $k$-th device, $L_k$, $\ell_k$, and $c_k$ are used to represent its total number of bits to be processed, its computation off-loading volume in terms of the number of bits, and the number of CPU cycles required to process a single bit, respectively. As for the local computing, upon denoting the computational capability at the $k$-th device in terms of the number of CPU cycles per second by $f^l_k$, the time required for carrying out the local computation is formulated as $D^l_k(\ell_k) = (L_k-\ell_k) c_k/ f^l_k$.
\item \emph{Edge computing:} we denote the maximum number of executable CPU cycles at the edge and the computational capability allocated to the $k$-th device by $f^e_{\text{total}}$ and $f^e_k$, respectively, which obey $\sum^K_{k=1} f^e_k \leq f^e_{\text{total}}$.
Here, it is assumed that the edge computing for the $k$-th device only begins its operation, when all its $\ell_k$ bits are completely off-loaded.
In this case, the total latency of edge computing is jointly constituted by the computation off-loading, and by the edge computing, as well as by the end-to-end delay of sending the computational result back.
Given that the computation result is typically of a small size \cite{mao2017survey}, the feedback latency can be negligible, upon using the technique of ultra-reliable low-latency communications \cite{bennis2018ultrareliable}.
Then, the total latency imposed by the computation off-loading and the edge computing is given by
$ D^e_k(\pmb{w}_k,\pmb{\theta},\ell_k,f^e_k) = \ell_k/R_k(\pmb{w}_k,\pmb{\theta}) + \ell_k c_k/f^e_k$.
\end{itemize}
To this end, the latency of the $k$-th device can be readily calculated by selecting the maximum value between those imposed by the local and by the edge computing, formulated as
\begin{IEEEeqnarray}{rCl}\label{eqn:latency}
D_k (\pmb{w}_k,\pmb{\theta},\ell_k, f^e_k)
& = &  \max \big\{ D^l_k(\ell_k), D^e_k(\pmb{w}_k,\pmb{\theta},\ell_k, f^e_k) \big\}  \\
& = &  \max \bigg\{ \frac{(L_k-\ell_k) c_k}{f^l_k}, \frac{\ell_k}{R_k(\pmb{w}_k,\pmb{\theta})} + \frac{\ell_k c_k}{f^e_k} \bigg\}. \nonumber
\end{IEEEeqnarray}

\subsection{Problem Formulation}
\label{sec: Problem Formulation}
In this paper, we aim for minimizing the weighted computational latency of all the devices, by jointly optimizing the computation off-loading volume $\pmb{\ell} = [\ell_1, \ell_2, \ldots, \ell_K]^T$, the edge computing resources $\pmb{f}^e = [f^e_1, f^e_2, \ldots, f^e_K]^T$ allocated to each device,  the MUD matrix $\pmb{W}$, and the IRS phase shift $\pmb{\theta}$.
Specifically, the weighted delay minimization problem is formulated as
\begin{subequations}
\begin{align}
\mathcal{P}0: & \mathop {\min} \limits_{\pmb{W},\pmb{\theta},\pmb{\ell},\pmb{f}^e}  \sum^{K}_{k=1} \varpi_k D_k(\pmb{w}_k,\pmb{\theta},\ell_k, f^e_k) \nonumber \\
& \text{s.t.} \quad 0 \leq \theta_n < 2\pi, \quad n = 1,2,\ldots,N, \label{eqn:P0_constraint1} \\
& \quad \quad \ell_k \in \{0,1,\ldots,L_k\}, \quad k = 1,2,\ldots,K, \label{eqn:P0_constraint2} \\
& \quad \quad \sum^K_{k=1} f^e_k \leq f^e_{\text{total}}, \label{eqn:P0_constraint3} \\
& \quad \quad f^e_k \geq 0, \quad k = 1,2,\ldots,K. \label{eqn:P0_constraint4}
\end{align}
\end{subequations}
where $\varpi_k$ represents the weight of the $k$-th device. \eqref{eqn:P0_constraint1} specifies the range of the phase shift of the IRS elements; \eqref{eqn:P0_constraint2} indicates that the computation off-loading volume should be an integer between $0$ and $L_k$ for the $k$-th device; Finally, \eqref{eqn:P0_constraint3} and \eqref{eqn:P0_constraint4} restrict the range of the edge computing resources allocated to each device.

\begin{remark}
In Problem~$\mathcal{P}0$, we have a total of four optimization variables, namely, the off-loading volume, edge computing resource allocation, MUD matrix, and IRS phase shift. The optimization of the former two variables is related to the computing setting, while the optimization of the other two specifies the communications design.
The difficulties of solving Problem $\mathcal{P}0$ are owing to three aspects. The first one is the segmented form of the OF. The second one is the coupling effect between the MUD matrix $\pmb{W}$ and the IRS phase shift vector $\pmb{\theta}$. The final one is that the OF is non-convex regarding the phase shift $\pmb{\theta}$. Hence, it is an open challenge to obtain a globally optimal solution directly. In this paper, a locally optimal solution is provided. Specifically, upon using the popular BCD technique for decoupling the communications and computing designs, the segmented form of the OF can be easily transformed to a more tractable form.
Similarly, optimal solutions can be provided for the MUD matrix $\pmb{W}$ and for the IRS phase shift vector $\pmb{\theta}$, after they are decoupled using the BCD technique in the communications design. To tackle the non-convexity regarding $\pmb{\theta}$, the Majorization-Minimization (MM) algorithm is invoked, which is capable of iteratively approaching a locally optimal solution at low complexity.
\end{remark}

\section{Joint Optimization of Computing and Communications Setting}
\label{sec:Solution to Problem}

The joint optimization of computing and computations settings is realized relying on the BCD technique. The pivotal idea of the BCD technique is to optimize one of the variables while fixing the other variables in an alternating manner, until the convergence of the OF is achieved.
In the rest of this section, the joint optimization of the off-loading volume and of the edge computing resource allocation is presented while fixing the communications setting, followed by the joint optimization of the MUD matrix and of the IRS phase shift while fixing the computing setting. Our goal is the joint optimization both of the communications and of the computing design.

\subsection{Joint Optimization of the Off-loading Volume and the Edge Computing Resource Allocation While Fixing the Communications Settings}

Given an MUD matrix $\pmb{W}$ and an IRS phase shift vector $\pmb{\theta}$, Problem $\mathcal{P}0$ can be simplified to
\begin{subequations}
\begin{align}
\mathcal{P}1: & \mathop {\min} \limits_{\pmb{\ell},\pmb{f}^e}  \sum^{K}_{k=1} \varpi_k D_k(\ell_k, f^e_k) \nonumber \\
& \text{s.t.}  \quad \eqref{eqn:P0_constraint2}, \eqref{eqn:P0_constraint3}, \eqref{eqn:P0_constraint4}.
\end{align}
\end{subequations}
The optimization of $\pmb{\ell}$ and $\pmb{f}^e$ can be decoupled, relying on the aforementioned BCD technique, detailed as follows.

\subsubsection{Optimization of $\pmb{\ell}$} The value of $\pmb{\ell}$ can be optimized, with the aid of the proposition below.
\begin{proposition}\label{proposition:offloading_volume}
Given an MUD matrix $\pmb{W}$, and an IRS phase shift coefficient vector $\pmb{\theta}$. as well as an edge computing resource allocation vector $\pmb{f}^e$, the optimal number of off-loaded bits is given by
\begin{IEEEeqnarray}{rCl}\label{eqn:offloading_volume_calculation}
\ell^*_k = \mathop{\arg \min} \limits_{\hat{\ell}_k \in \big\{ \lfloor \hat{\ell}^*_k \rfloor, \lceil \hat{\ell}^*_k\rceil \big\} } D_k (\hat{\ell}_k),
\end{IEEEeqnarray}
where $\lfloor \cdot \rfloor$ and $\lceil \cdot \rceil$ represent the floor and ceiling operations, respectively, and $\hat{\ell}^*_k$ is selected for ensuring that the value of $ D^l_k(\hat{\ell}_k)$ becomes equivalent to that of $D^e_k(\hat{\ell}_k)$, i.e.
\begin{IEEEeqnarray}{rCl}\label{eqn:offloading_volume_relax}
\hat{\ell}^*_k = \frac{L_k c_k R_k f^e_k}{f^e_k f^l_k + c_kR_k \big(f^e_k + f^l_k \big)}.
\end{IEEEeqnarray}
\end{proposition}

\textbf{\textit{Proof: }} See Appendix~\ref{app:Proposition_1}. \hspace{4.5cm} $\blacksquare$

\begin{algorithm}[t]\small
\caption{Joint optimization of $\pmb{\ell}$ and $\pmb{f}^e$, given $\pmb{W}$ and $\pmb{\theta}$}
\begin{algorithmic}
 \renewcommand{\algorithmicrequire}{\textbf{Input:}}
 \renewcommand{\algorithmicensure}{\textbf{Output:}}
 \REQUIRE $\pmb{h}_{r,k}$, $\pmb{h}_{d,k}$, $\pmb{G}$, $B$, $p_t$, $\sigma^2$, $K$, $\varpi_k$, $L_k$, $c_k$, $f^l_k$, $f^e_{\text{total}}$, $t^{\text{max}}_1$, $\epsilon$, $\pmb{W}$, $\pmb{\theta}$, and ${\hat{\pmb{f}}^e}$ satisfying \eqref{eqn:P0_constraint3} and \eqref{eqn:P0_constraint4}
 \ENSURE  Optimal ${\pmb{\ell}}^*$ and ${\pmb{f}^e}^*$, given $\pmb{W}$ and $\pmb{\theta}$
\\ \textbf{1. Initialization}
\STATE initialize $t_1 = 0$, $\epsilon_1^{(0)} = 1$, ${\pmb{f}^e}^{(0)} \leftarrow \hat{\pmb{f}}^e$
\STATE calculate $\pmb{R}$ using \eqref{eqt:throughput}
\\ \textbf{2. Joint optimization of $\pmb{\ell}$ and $\pmb{f}^e$}
\WHILE{$\epsilon_1^{(t_1)} > \epsilon$ $\&\&$ $t_1 < t^{\text{max}}_1$}
\STATE $\bullet$ calculate $\pmb{\ell}^{(t_1+1)}$ using \eqref{eqn:offloading_volume_relax}
\STATE $\bullet$ calculate ${\pmb{f}^e}^{(t_1+1)}$ and $\mu$ by using \eqref{eqn:f^e_k} and the bisection search method, respectively
\STATE $\bullet$ $\epsilon_1^{(t_1+1)} =\frac{ \big|\text{obj}\big(\pmb{\ell}^{(t_1+1)},{\pmb{f}^e}^{(t_1+1)}\big)- \text{obj}\big(\pmb{\ell}^{(t_1)},{\pmb{f}^e}^{(t_1)}\big)\big|}{\text{obj}\big(\pmb{\ell}^{(t_1+1)},{\pmb{f}^e}^{(t_1+1)}\big)}$
\STATE $\bullet$ $t_1 \leftarrow t_1 + 1$
\ENDWHILE
\\ \textbf{3. Output optimal ${\pmb{\ell}}^*$ and ${\pmb{f}^e}^*$}
\STATE ${\pmb{\ell}}^* \leftarrow \pmb{\ell}^{(t_1)}$ and ${\pmb{f}^e}^* \leftarrow {\pmb{f}^e}^{(t_1)}$
\end{algorithmic}
\label{alg:joint_optimization1}
\end{algorithm}

\subsubsection{Optimization of $\pmb{f}^e$}
Here, the edge computing resource allocation $\pmb{f}^e$ is optimized, while fixing the MUD matrix $\pmb{W}$, the IRS phase shift coefficient vector $\pmb{\theta}$, and the off-loading volume $\pmb{\ell}$.
Specifically, upon substituting \eqref{eqn:offloading_volume_relax} into the OF of Problem $\mathcal{P}1$, the problem can be reformulated as:
\begin{subequations}
\begin{align}
\mathcal{P}1\text{-}E: & \mathop {\min} \limits_{\pmb{f}^e}  \sum^{K}_{k=1} \frac{\varpi_k (L_k c^2_k R_k + L_k c_k f^e_k )}{f^e_k f^l_k + c_kR_k (f^e_k + f^l_k )} \nonumber \\
& \text{s.t.} \quad \eqref{eqn:P0_constraint3}, \eqref{eqn:P0_constraint4}.
\end{align}
\end{subequations}
Problem $\mathcal{P}1\text{-}E$ can be proved to be a convex optimization problem following the proposition below.
\begin{proposition}\label{proposition:convex_optimization}
Problem $\mathcal{P}1\text{-}E$ is a convex optimization problem.
\end{proposition}

\textbf{\textit{Proof: }} See Appendix~\ref{app:Proposition_2}. \hspace{4.5cm} $\blacksquare$

Since Problem $\mathcal{P}1\text{-}E$ is convex and the Slater's condition \cite{boyd2004convex} is satisfied \footnote{In words, the Slater's condition for convex programming states that strong duality holds if all constraints are satisfied and the nonlinear constraints are satisfied with strict inequalities.}, the Karush–Kuhn–Tucker (KKT) may be imposed on the problem for finding its optimal solution.
Specifically, the Lagrangian function associated with Problem $\mathcal{P}1\text{-}E$ is given by
\begin{IEEEeqnarray}{rCl}
\mathcal{L} (\pmb{f}^e, \mu, \pmb{\nu})
& = & \sum^{K}_{k=1} \frac{\varpi_k (L_k c^2_k R_k + L_k c_k f^e_k )}{c_k R_k f^l_k + (f^l_k + c_k R_k) f^e_k} \nonumber \\
&& + \mu \bigg( \sum^K_{k=1} f^e_k - f^e_{\text{total}} \bigg),
\end{IEEEeqnarray}
where the variable $\mu$ is the non-negative Lagrange multiplier,
while the optimal edge computing resource allocation vector ${\pmb{f}^e}^*$ and the optimal Lagrange multiplier $\mu^*$ should satisfy the following KKT conditions, for $k = 1,2,\ldots, K$:
\begin{align}
& \frac{\partial \mathcal{L} }{\partial f^e_k} = \frac{ - \varpi_k L_k c^3_k R^2_k}{\big[c_k R_k f^l_k + (f^l_k + c_k R_k) {f^e_k}^* \big]^2} + \mu^* = 0, \label{eqn:KKT_1_E_1} \\
& \mu^* \bigg( \sum^K_{k=1} {f^e_k}^* - f^e_{\text{total}} \bigg) = 0, \label{eqn:KKT_1_E_2} \\
& {f^e_k}^* \geq 0. \label{eqn:KKT_1_E_3}
\end{align}
The value of $f^e_k$ can be directly derived from \eqref{eqn:KKT_1_E_1} for a given $\mu$, which is written as
\begin{IEEEeqnarray}{rCl}\label{eqn:f^e_k}
f^e_k = \frac{\sqrt{\frac{\varpi_k L_k c^3_k R^2_k}{\mu}}-c_k R_k f^l_k}{f^l_k + c_k R_k}, \quad k = 1,\ldots,K.
\end{IEEEeqnarray}
In order to ensure $f^e_k \geq 0$ in \eqref{eqn:f^e_k}, we have $\sqrt{\frac{\varpi_k L_k c^3_k R^2_k}{\mu}}-c_k R_k f^l_k \geq 0$, which is reformulated as $\mu \leq \frac{\varpi_k L_k c_k}{{f^l_k}^2}$. Given that $\mu \neq 0$ in \eqref{eqn:f^e_k}, the optimal $\mu^*$ can be found in the range of $(\mu_l, \mu_u] = \Big(0, \mathop {\min} \limits_{k} \big(\frac{\varpi_k L_k c_k}{{f^l_k}^2}\big) \Big]$ to ensure \eqref{eqn:KKT_1_E_2}, using the well-known bisection search method associated with the termination coefficient of $\epsilon$, because $\sum^K_{k=1} {f^e_k}$ can be proved to be monotonically decreasing with respect to $\mu$. The procedure of solving Problem $\mathcal{P}1$ is summarized in Algorithm~\ref{alg:joint_optimization1}. The complexity of Algorithm~\ref{alg:joint_optimization1} is dominated by calculating ${\pmb{f}^e}^{(t_1+1)}$ using \eqref{eqn:f^e_k} and by calculating $\mu$ using the bisection search method. Its complexity is on the order of $\mathcal{O}\big( \log_2 (\frac{\mu_u-\mu_l}{\epsilon}) K \big)$.
Thus the total complexity of Algorithm~\ref{alg:joint_optimization1} is $\mathcal{O}\big(t^{\text{max}}_1 \log_2 (\frac{\mu_u-\mu_l}{\epsilon})K\big)$.

\subsection{Joint Optimization of the MUD Matrix and the IRS Phase Shift Coefficient While Fixing the Computing Settings}
\label{sec:Joint Optimization of MUD Matrix and IRS Phase Shift Coefficient While Fixing Computing Settings}

Given an off-loading volume vector $\pmb{\ell}$ and an edge computing resource allocation vector $\pmb{f}^e$, Problem $\mathcal{P}0$ is reformulated as
\begin{subequations}
\begin{align}
\mathcal{P}2: & \mathop {\min} \limits_{\pmb{W},\pmb{\theta}}  \sum^{K}_{k=1} \varpi_k D_k(\pmb{w}_k,\pmb{\theta}) \nonumber \\
& \text{s.t.} \quad 0 \leq \theta_n < 2\pi, \quad n = 1,2,\ldots,N.
\end{align}
\end{subequations}

\begin{remark}
The challenges of solving Problem $\mathcal{P}2$ are due to two aspects. The first one is the segmented form of $D_k(\pmb{w}_k,\pmb{\theta})$ that is caused by the operation $\max$ as detailed in \eqref{eqn:latency}, while the second issue is that the OF is the summation of fractional functions, with respect to $\pmb{W}$ and $\pmb{\theta}$ as shown in the OF of Problem~$\mathcal{P}2\text{-}E1$ below, which makes the problem a non-convex sum-of-ratios optimization. In order to tackle these two issues, we transform the problem as follows.
\end{remark}

\subsubsection{Problem Transformation}
As detailed in Proposition~\ref{proposition:offloading_volume}, the optimal solution of Problem $\mathcal{P}0$ results in $D_k = D^l_k = D^e_k$. Hence upon replacing $D_k$ by $D^e_k$ and removing the constant terms, Problem $\mathcal{P}2$ is reformulated as:
\begin{subequations}
\begin{align}
\mathcal{P}2\text{-}E1: & \mathop {\min} \limits_{\pmb{W},\pmb{\theta}}  \sum^{K}_{k=1}  \frac{\varpi_k \ell_k}{R_k(\pmb{w}_k,\pmb{\theta}) } \nonumber \\
& \text{s.t.} \quad 0 \leq \theta_n < 2\pi, \quad n = 1,2,\ldots,N.
\end{align}
\end{subequations}
It is then rewritten as the following equivalent form:
\begin{subequations}
\begin{align}
\mathcal{P}2\text{-}E2: & \mathop {\min} \limits_{\pmb{W},\pmb{\theta},\pmb{\beta}}  \sum^{K}_{k=1} \beta_k  \nonumber \\
& \text{s.t.} \quad \frac{\varpi_k \ell_k}{R_k(\pmb{w}_k,\pmb{\theta})} \leq \beta_k, \quad k = 1,2,\ldots,K, \\
& \quad \quad 0 \leq \theta_n < 2\pi, \quad n = 1,2,\ldots,N.
\end{align}
\end{subequations}
The following proposition may assist us in solving Problem~$\mathcal{P}2\text{-}E2$.

\begin{algorithm}[t!]\small
\caption{Joint optimization of $\pmb{W}$ and $\pmb{\theta}$, given $\pmb{\ell}$ and $\pmb{f}^e$}
\begin{algorithmic}
 \renewcommand{\algorithmicrequire}{\textbf{Input:}}
 \renewcommand{\algorithmicensure}{\textbf{Output:}}
 \REQUIRE $\pmb{h}_{r,k}$, $\pmb{h}_{d,k}$, $\pmb{G}$, $B$, $p_t$, $\sigma^2$, $\varpi_k$, $\pmb{\ell}$, and $\pmb{f}^e$
 \ENSURE  Optimal $\pmb{W}^*$ and $\pmb{\theta}^*$, given $\pmb{\ell}$ and $\pmb{f}^e$
\\ \textbf{1. Initialization}
\STATE initialize $t_2 = 0$, $\zeta \in (0,1)$, $\epsilon \in (0,1)$, and $\pmb{\theta}^{(0)}$ satisfying \eqref{eqn:P0_constraint1}
\STATE  calculate $\pmb{W}^{(0)}$ and $\pmb{R}^{(0)}$ using \eqref{eqn:MMSE_MUD} and \eqref{eqn:MMSE_rate}, respectively
\STATE  calculate $\pmb{\lambda}^{(0)}$ and $\pmb{\beta}^{(0)}$ using \eqref{eqt:variable_relationship}
\\ \textbf{2. Joint optimization of $\pmb{W}$, $\pmb{\theta}$, $\pmb{\lambda}$ and $\pmb{\beta}$}
\REPEAT
\STATE $\bullet$ update $\pmb{W}^{(t_2+1)}$ and $\pmb{\theta}^{(t_2+1)}$ using Algorithm~\ref{alg:joint_optimization2}
\STATE $\bullet$ update $\pmb{\lambda}^{(t_2+1)}$ and $\pmb{\beta}^{(t_2+1)}$ as follows
\begin{equation}
\lambda_k^{(t_2+1)} = \lambda_k^{(t_2)} - \frac{\zeta^{i^{(t_2+1)}} \chi_k\big(\lambda_k^{(t_2)}\big)}{R_k\big(\pmb{w}_k^{(t_2+1)},\pmb{\theta}^{(t_2+1)}\big)},
\end{equation}
and
\begin{equation}
\beta^{(t_2+1)} = \beta^{(t_2)} - \frac{\zeta^{i^{(t_2+1)}} \kappa_k\big(\beta_k^{(t_2)}\big)}{R_k\big(\pmb{w}_k^{(t_2+1)},\pmb{\theta}^{(t_2+1)}\big)},
\end{equation}
where $i^{(t_2+1)}$ is the smallest integer among $i \in \{1, 2, 3, \ldots \}$ satisfying
\begin{IEEEeqnarray}{rCl}
&& \sum^K_{k=1} \bigg| \chi_k \bigg( \lambda_k^{(t_2)} - \frac{\zeta^i \chi_k(\lambda_k^{(t_2)})}{R_k(\pmb{w}_k^{(t_2+1)},\pmb{\theta}^{(t_2+1)})} \bigg) \bigg|^2 \nonumber \\
&& \quad + \sum^K_{k=1} \bigg| \kappa_k \bigg( \beta^{(t_2)} - \frac{\zeta^i \kappa_k(\beta^{(t_2)})}{R_k(\pmb{w}_k^{(t_2+1)},\pmb{\theta}^{(t_2+1)})} \bigg) \bigg|^2 \quad \quad \nonumber \\
&& \quad \leq (1 - \epsilon_3 \zeta^i)^2 \sum^K_{k=1} \Big[ \big| \chi_k\big(\lambda_k^{(t_2)}\big) \big|^2 + \big| \kappa_k\big(\beta^{(t_2)}\big) \big|^2 \Big].
\end{IEEEeqnarray}
\STATE $\bullet$ $t_2 \leftarrow t_2 +1$
\UNTIL the following conditions are achieved
\begin{IEEEeqnarray}{rCl}
\lambda_k^{(t_2)} R_k(\pmb{w}_k^{(t_2)},\pmb{\theta}^{(t_2)}) - 1 = 0,
\end{IEEEeqnarray}
\begin{IEEEeqnarray}{rCl}
\beta_k R_k(\pmb{w}_k^{(t_2)},\pmb{\theta}^{(t_2)}) - \varpi_k \ell_k= 0
\end{IEEEeqnarray}
\\ \textbf{3. Output optimal ${\pmb{W}}^*$ and ${\pmb{\theta}}^*$}
\STATE ${\pmb{W}}^* \leftarrow \pmb{W}^{(t_2)}$ and ${\pmb{\theta}}^* \leftarrow {\pmb{\theta}}^{(t_2)}$
\end{algorithmic}
\label{alg:joint_optimization3}
\end{algorithm}

\begin{proposition}\label{Proposition:problem_transfer}
If $(\pmb{W}^*,\pmb{\theta}^*,\pmb{\beta}^*)$ is the solution of Problem $\mathcal{P}2\text{-}E2$, a $\pmb{\lambda}^* = [ \lambda_1, \lambda_2, \ldots , \lambda_K]$ exists that $(\pmb{W}^*,\pmb{\theta}^*)$ satisfies the KKT conditions of the following problem, when we set $\pmb{\beta} = \pmb{\beta}^*$ and $\pmb{\lambda} = \pmb{\lambda}^*$
\begin{subequations}
\begin{align}
\mathcal{P}2\text{-}E3: & \mathop {\min} \limits_{\pmb{W},\pmb{\theta}}  \sum^{K}_{k=1}
\lambda_k \big[  \varpi_k \ell_k -\beta_k R_k(\pmb{w}_k,\pmb{\theta}) \big] \nonumber \\
& \text{s.t.}  \quad 0 \leq \theta_n < 2\pi, \quad  n = 1,2,\ldots,N. \label{eqn:P2-E3_constraint}
\end{align}
\end{subequations}
Furthermore, $(\pmb{W}^*,\pmb{\theta}^*)$ also satisfies the following equations, when we set $\pmb{\beta} = \pmb{\beta}^*$ and $\pmb{\lambda} = \pmb{\lambda}^*$
\begin{equation}\label{eqt:variable_relationship}
\begin{cases}
\lambda_k = \frac{1}{R_k(\pmb{w}_k^*,\pmb{\theta}^*)}, \quad  k = 1,2,\ldots,K,\\
\beta_k = \frac{\varpi_k \ell_k}{R_k(\pmb{w}_k^*,\pmb{\theta}^*)}, \quad  k = 1,2,\ldots,K.
\end{cases}
\end{equation}
Correspondingly, if $(\pmb{W}^*,\pmb{\theta}^*)$ is a solution to Problem $\mathcal{P}2\text{-}E3$ and satisfies \eqref{eqt:variable_relationship} when we set $\pmb{\beta} = \pmb{\beta}^*$ and $\pmb{\lambda} = \pmb{\lambda}^*$, $(\pmb{W}^*,\pmb{\theta}^*,\pmb{\beta}^*)$ is the solution of Problem $\mathcal{P}2\text{-}E2$ associated with the Lagrange multiplier $\pmb{\lambda} = \pmb{\lambda}^*$.
\end{proposition}

\textbf{\textit{Proof: }} See Appendix~\ref{app:Proposition_3}. \hspace{4.5cm} $\blacksquare$

To this end, the sum-of-ratios form in Problem $\mathcal{P}2\text{-}E1$ has been transformed to a parameterized subtractable form in Problem $\mathcal{P}2\text{-}E3$, which can be solved in two steps \cite{jong2012efficient,he2013coordinated,pan2019caching}: the first step is to obtain $\pmb{W}^*$ and $\pmb{\theta}^*$ by solving Problem $\mathcal{P}2\text{-}E3$, given $\pmb{\beta}$ and $\pmb{\lambda}$; the second step is to update $\pmb{\beta}$ and $\pmb{\lambda}$ using the modified Newton's method until the convergence is achieved. The procedure is summarized in Algorithm~2, where we have
\begin{IEEEeqnarray}{rCl}
\chi_k(\lambda_k) = \lambda_k R_k(\pmb{w}_k^*,\pmb{\theta}^*) - 1, \quad k = 1,2,\ldots,K,
\end{IEEEeqnarray}
\begin{IEEEeqnarray}{rCl}
\kappa_k( \beta_k) = \beta_k R_k(\pmb{w}_k^*,\pmb{\theta}^*) - \varpi_k \ell_k, \quad k = 1,2,\ldots,K.
\end{IEEEeqnarray}
The complexity of Algorithm~2 is analyzed at the end of Section~\ref{sec:Joint Optimization of MUD Matrix and IRS Phase Shift Coefficient While Fixing Computing Settings}.

Let us now focus our attention on the first step of solving Problem~$\mathcal{P}2\text{-}E3$, i.e. optimizing $\pmb{W}^*$ and $\pmb{\theta}^*$, given a set of $\pmb{\beta}$ and $\pmb{\lambda}$ as well as an off-loading volume vector $\pmb{\ell}$. In this case, Problem $\mathcal{P}2\text{-}E3$ can be simplified to $\mathop {\max} \limits_{\pmb{W},\pmb{\theta}}  \sum^{K}_{k=1} \lambda_k \beta_k R_k(\pmb{w}_k,\pmb{\theta})$ subject to \eqref{eqn:P2-E3_constraint}, which constitutes a weighted sum-rate maximization problem. As revealed in \cite{shi2011iteratively}, maximizing the weighted sum-rate can be accomplished via weighted MSE minimization. The latter problem is easier to handle, because it is convex regarding each optimization variable, while fixing others. As such, we focus our attention on constructing the corresponding weighted MSE minimization problem. Specifically, following Theorem \RN{1} in \cite{shi2011iteratively}, we introduce an auxiliary weight variable $\Upsilon_k$ for the $k$-th device and formulate the corresponding weighted MSE minimization problem as:
\begin{subequations}
\begin{align}
\mathcal{P}2\text{-}E4: & \mathop {\min} \limits_{\pmb{W},\pmb{\theta}}  \sum^{K}_{k=1}
\big[ \Upsilon_k e_k (\pmb{W},\pmb{\theta}) \nonumber \\
&  \qquad  -  \lambda_k \beta_k \log_2 (\lambda_k^{-1} \beta_k^{-1} \Upsilon_k) - \lambda_k \beta_k \big] \nonumber \\
& \text{s.t.}  \quad 0 \leq \theta_n < 2\pi, \quad \forall n \in \{1,2,\ldots,N \},
\end{align}
\end{subequations}
where the mathematical expression of $\{\Upsilon_k\}$ is given in Section~\ref{sec:Auxiliary Variable Design} and $e_k$ represents the MSE of the $k$-th user, which is given by
\begin{IEEEeqnarray}{rCl}\label{eqn:mse}
e_k(\pmb{W},\pmb{\theta})
& \triangleq &  \mathbb{E} \big[(\hat{s}_k - s_k)(\hat{s}_k - s_k)^H \big] \nonumber \\
&  = & \big[ \sqrt{p_t} \pmb{w}^H_k (\pmb{h}_{d,k} + \pmb{G} \pmb{\Theta} \pmb{h}_{r,k} )-1\big] \times \nonumber \\
&&  \big[ \sqrt{p_t} \pmb{w}^H_k (\pmb{h}_{d,k} + \pmb{G} \pmb{\Theta} \pmb{h}_{r,k} )-1\big]^H \nonumber \\
& & + p_t \sum^K_{j\neq k} \pmb{w}_k^H (\pmb{h}_{d,j} + \pmb{G} \pmb{\Theta} \pmb{h}_{r,j} )(\pmb{h}_{d,j} + \pmb{G} \pmb{\Theta} \pmb{h}_{r,j} )^H \pmb{w}_k \nonumber \\
&& + \sigma^2 \pmb{w}^H_k \pmb{w}_k.
\end{IEEEeqnarray}
As such, compared to Problem $\mathcal{P}2\text{-}E3$, Problem $\mathcal{P}2\text{-}E4$ becomes more tractable, because given an IRS phase shift coefficient vector, the OF of Problem $\mathcal{P}2\text{-}E4$ is convex regarding an optimization variable, while fixing the other one.
Again, the BCD technique is invoked for solving this problem as follows.

\subsubsection{MUD Matrix Design}
In Problem $\mathcal{P}2\text{-}E4$, fixing the phase shift coefficient vector $\pmb{\theta}$ and the auxiliary variable $\Upsilon_k$, the MUD vector can be obtained by forcing the first-order derivative of the OF with respect to $\pmb{w}_k$ as $0$. After several steps of mathematical manipulations, it is readily observed that the above minimization is equivalent to minimizing the weighted MSE. Then, the MUD vector is given by \cite[Sec.~6.2.3]{yang2009multicarrier}
\begin{IEEEeqnarray}{rCl}\label{eqn:MMSE_MUD}
\pmb{w}_k = \sqrt{p_t} \pmb{J}^{-1} (\pmb{h}_{d,k} + \pmb{G} \pmb{\Theta} \pmb{h}_{r,k} ),
\end{IEEEeqnarray}
where $\pmb{J} = p_t \sum^K_{j=1} (\pmb{h}_{d,j} + \pmb{G} \pmb{\Theta} \pmb{h}_{r,j} )(\pmb{h}_{d,j} + \pmb{G} \pmb{\Theta} \pmb{h}_{r,j} )^H + \sigma^2 \pmb{I}_M$.

\subsubsection{Auxiliary Variable Design}
\label{sec:Auxiliary Variable Design}
Fixing $\pmb{\theta}$ and $\pmb{w}_k$, the optimal auxiliary variable can be obtained by minimizing the OF of Problem $\mathcal{P}2\text{-}E4$ with respect to $\Upsilon_k$, given by
\begin{IEEEeqnarray}{rCl}\label{eqn:auxiliary_variable}
\Upsilon_k = \lambda_k \beta_k ( e_k )^{-1}.
\end{IEEEeqnarray}
Furthermore, substituting \eqref{eqn:MMSE_MUD} into \eqref{eqn:mse}, the MSE becomes
\begin{IEEEeqnarray}{rCl}\label{eqn:MMSE_MSE}
e^{\text{MMSE}}_{k} = 1 - p_t (\pmb{h}_{d,k} + \pmb{G} \pmb{\Theta} \pmb{h}_{r,k} )^H \pmb{J}^{-1} (\pmb{h}_{d,k} + \pmb{G} \pmb{\Theta} \pmb{h}_{r,k} ). \nonumber \\
\end{IEEEeqnarray}
Bearing in mind that the relationship between the SINR and the MSE of the system equipped with the minimum mean square error (MMSE) MUD is given by $\gamma_k = (e^{\text{MMSE}}_{k} )^{-1} -1$ \cite[Sec.~6.2.3]{yang2009multicarrier}, \eqref{eqt:throughput} may be reformulated as
\begin{IEEEeqnarray}{rCl}\label{eqn:MMSE_rate}
R_k = - B \log_2 \big(e^{\text{MMSE}}_{k} \big).
\end{IEEEeqnarray}

\subsubsection{IRS Phase Shift Coefficient Design}
\label{sec: IRS Phase Shift Coefficient Design}
In this subsection, we focus our attention on optimizing the reflection phase shift coefficients $\pmb{\theta}$, while fixing the auxiliary variable $\Upsilon_k$ and the MUD matrix $\pmb{W}$. Specifically, by substituting \eqref{eqn:mse} into the OF of Problem $\mathcal{P}2\text{-}E4$ and removing the terms that are independent of the phase shift coefficient vector $\pmb{\theta}$, Problem $\mathcal{P}2\text{-}E4$ is reformulated as:
\begin{align}
\mathcal{P}2\text{-}E5: & \mathop {\min} \limits_{\pmb{\theta}} \sum^{K}_{k=1} \sum^{K}_{j=1} \Upsilon_k p_t \pmb{w}^H_k \pmb{h}_j \pmb{h}_j^H \pmb{w}_k \nonumber \\
& \quad - \sum^{K}_{k=1} \Upsilon_k \sqrt{p_t} \pmb{h}_k^H \pmb{w}_k - \sum^{K}_{k=1} \Upsilon_k \sqrt{p_t} \pmb{w}_k^H \pmb{h}_k  \nonumber \\
& \text{s.t.}  \quad 0 \leq \theta_n \leq 2\pi, \quad \forall n \in \{1,2,\ldots,N \},
\end{align}
where the first and the second terms in the OF can be respectively formulated in expansion forms as
\begin{IEEEeqnarray}{l}
\quad \sum^{K}_{k=1} \sum^{K}_{j=1} \Upsilon_k p_t \pmb{w}^H_k \pmb{h}_j \pmb{h}_j^H \pmb{w}_k  \nonumber \\
 =   \sum^{K}_{k=1} \sum^{K}_{j=1} \big(  \Upsilon_k p_t \pmb{w}_k^H \pmb{G} \pmb{\Theta} \pmb{h}_{r,j} \pmb{h}_{r,j}^H \pmb{\Theta}^H \pmb{G}^H \pmb{w}_k \nonumber \\
\quad + \Upsilon_k p_t \pmb{w}_k^H \pmb{h}_{d,j} \pmb{h}_{r,j}^H \pmb{\Theta}^H \pmb{G}^H \pmb{w}_k  \nonumber \\
\quad + \Upsilon_k p_t \pmb{w}_k^H \pmb{G} \pmb{\Theta} \pmb{h}_{r,j} \pmb{h}_{d,j}^H \pmb{w}_k + \Upsilon_k p_t \pmb{w}_k^H \pmb{h}_{d,j} \pmb{h}_{d,j}^H \pmb{w}_k \big),
\end{IEEEeqnarray}
and
\begin{IEEEeqnarray}{l}
\quad \sum^{K}_{k=1} \Upsilon_k \sqrt{p_t} \pmb{h}_k^H \pmb{w}_k \nonumber \\
= \sum^{K}_{k=1} \big( \Upsilon_k \sqrt{p_t} \pmb{h}_{d,k}^H \pmb{w}_k +  \Upsilon_k \sqrt{p_t} \pmb{h}_{r,k}^H \pmb{\Theta}^H  \pmb{G}^H \pmb{w}_k \big).
\end{IEEEeqnarray}
Upon defining $\pmb{A} \triangleq \sum^{K}_{k=1} \Upsilon_k p_t \pmb{G}^H \pmb{w}_k \pmb{w}_k^H \pmb{G}$, $\pmb{B} \triangleq \sum^{K}_{j=1} \pmb{h}_{r,j} \pmb{h}_{r,j}^H$, $\pmb{C} \triangleq \sum^{K}_{k=1} \sum^{K}_{j=1} \Upsilon_k p_t \pmb{h}_{r,j} \pmb{h}_{d,j}^H  \pmb{w}_k \pmb{w}_k^H \pmb{G}$, and $\pmb{D} \triangleq \sum^{K}_{k=1} \Upsilon_k \sqrt{p_t} \pmb{h}_{r,k} \pmb{w}_k^H \pmb{G} $, Problem $\mathcal{P}2\text{-}E5$ may be rewritten as:
\begin{align}
\mathcal{P}2\text{-}E6: & \mathop {\min} \limits_{\pmb{\theta}} \text{tr}(\pmb{\Theta}^H \pmb{A} \pmb{\Theta} \pmb{B}) + \text{tr} \big[ \pmb{\Theta}^H (\pmb{C} - \pmb{D})^H \big] \nonumber \\
& \quad +  \text{tr} \big[\pmb{\Theta} (\pmb{C} - \pmb{D})  \big] \nonumber \\
& \text{s.t.}  \quad 0 \leq \theta_n \leq 2\pi, \quad \forall n \in \{1,2,\ldots,N \}.
\end{align}
Defining $\pmb{\phi} \triangleq [\phi_1, \ldots, \phi_N]^T$ where $\phi_n = e^{j\theta_n}$, and $\pmb{v} = \big[ [\pmb{C} - \pmb{D}]_{1,1},\ldots, [\pmb{C} - \pmb{D}]_{N,N} \big]^T$, we have
\begin{IEEEeqnarray}{rCl}
\text{tr}(\pmb{\Theta}^H \pmb{A} \pmb{\Theta} \pmb{B}) = \pmb{\phi}^H ( \pmb{A} \odot \pmb{B}) \pmb{\phi},
\end{IEEEeqnarray}
where $\odot$ represents the Hadamard product, and
\begin{IEEEeqnarray}{rCl}
\text{tr} \big[ \pmb{\Theta}^H (\pmb{C} - \pmb{D})^H \big] = \pmb{v}^H \pmb{\phi}^*, \text{ }
\text{tr} \big[ \pmb{\Theta} (\pmb{C} - \pmb{D}) \big] = \pmb{\phi}^T \pmb{v}.
\end{IEEEeqnarray}
Further defining $\pmb{\Psi} \triangleq \pmb{A} \odot \pmb{B}$, we may equivalently rewrite Problem $\mathcal{P}2\text{-}E6$ as:
\begin{align}
\mathcal{P}2\text{-}E7: & \mathop {\min} \limits_{\pmb{\phi}} f(\pmb{\phi}) = \pmb{\phi}^H \pmb{\Psi} \pmb{\phi} + 2 \Re \big\{ \pmb{\phi}^H \pmb{v}^* \big\} \nonumber \\
& \text{s.t.}  \quad  |\phi_n | = 1, \quad \forall n \in \{1,2,\ldots,N \} \label{eqn:unit_modulus_constraint}.
\end{align}
Problem $\mathcal{P}2\text{-}E7$ is a non-convex one because of the unit modulus constraint on $\phi_n$. In the following, the MM algorithm \cite{sun2016majorization} is invoked for solving this problem, which has two steps. In the majorization step, we construct a continuous surrogate function $g(\pmb{\phi}|\pmb{\phi}^t)$, which represents the upperbound of $f(\pmb{\phi})$. Then in the minimization step, $\pmb{\phi}$ is updated by $\pmb{\phi}^{t+1} \in \mathop {\arg \min} \limits_{\pmb{\phi}} g(\pmb{\phi}|\pmb{\phi}^t)$.
As such, we may initialize $\pmb{\phi}^0$ that satisfies the constraint \eqref{eqn:unit_modulus_constraint}, and then use the MM algorithm to generate a sequence of feasible vectors $\{\pmb{\phi}^t \}$, where $t$ refers to the iteration index.
Now the surrogate function is constructed with the aid of the proposition below.

\begin{algorithm}[t!]\small
\caption{Joint optimization of $\pmb{W}$ and $\pmb{\theta}$, given $\pmb{\lambda}$ and $\pmb{\beta}$}
\begin{algorithmic}
 \renewcommand{\algorithmicrequire}{\textbf{Input:}}
 \renewcommand{\algorithmicensure}{\textbf{Output:}}
 \REQUIRE $\pmb{h}_{r,k}$, $\pmb{h}_{d,k}$, $\pmb{G}$, $B$, $p_t$, $\sigma^2$, $\varpi_k$, $t^{\text{max}}_2$, $\epsilon$, $\pmb{\lambda}$ and $\pmb{\beta}$
 \ENSURE  Optimal $\pmb{W}^*$ and $\pmb{\theta}^*$, given $\pmb{\lambda}$ and $\pmb{\beta}$
\\ \textbf{1. Initialization}
\STATE initialize $t_3 = 0$, $\epsilon_3^{(0)} = 1$, $\pmb{\theta}^{(0)}$ satisfying \eqref{eqn:P0_constraint1}
\\ \textbf{2. Joint optimization of $\pmb{W}$ and $\pmb{\theta}$}
\WHILE{$\epsilon_3^{(t_2)} > \epsilon$ $\&\&$ $t_3 < t^{\text{max}}_3$}
\STATE $\bullet$ calculate $\pmb{W}^{(t_3+1)}$ using \eqref{eqn:MMSE_MUD}
\STATE $\bullet$ calculate $\pmb{\Upsilon}^{(t_3+1)}$ using \eqref{eqn:auxiliary_variable}
\STATE $\bullet$ calculate ${\pmb{\theta}}^{(t_3+1)}$ by solving Problem $\mathcal{P}2\text{-}E6$ with the aid of the MM algorithm
\STATE $\bullet$ $\epsilon_3^{(t_3+1)} =\frac{ \big|\text{obj}\big(\pmb{W}^{(t_3+1)},{\pmb{\theta}}^{(t_3+1)}\big)- \text{obj}\big(\pmb{W}^{(t_3)},{\pmb{\theta}}^{(t_3)}\big)\big|}{\text{obj}\big(\pmb{W}^{(t_3+1)},{\pmb{\theta}}^{(t_3+1)}\big)}$
\STATE $\bullet$ $t_3 \leftarrow t_3 +1$
\ENDWHILE
\\ \textbf{3. Output optimal ${\pmb{W}}^*$ and ${\pmb{\theta}}^*$, given $\pmb{\lambda}$ and $\pmb{\beta}$}
\STATE ${\pmb{W}}^* \leftarrow \pmb{W}^{(t_3)}$ and ${\pmb{\theta}}^* \leftarrow {\pmb{\theta}}^{(t_3)}$
\end{algorithmic}
\label{alg:joint_optimization2}
\end{algorithm}

\begin{proposition}
Denoting the maximum eigenvalue of $\pmb{\Psi}$ by $\hat{\lambda}_{max}$ and given a solution $\pmb{\phi}^t$ at the $t$-th iteration, we have the inequality below
\begin{IEEEeqnarray}{rCl}\label{eqn:surrogate_function}
f(\pmb{\phi}) & \leq & \pmb{\phi}^H \hat{\lambda}_{max} \pmb{I}_N \pmb{\phi} - 2 \Re \big\{ \pmb{\phi}^H (\hat{\lambda}_{max} \pmb{I}_N - \pmb{\Psi}) \pmb{\phi}^t \big\} \nonumber \\
& &+ (\pmb{\phi}^t)^H (\hat{\lambda}_{max} \pmb{I}_N - \pmb{\Psi}) \pmb{\phi}^t + 2 \Re \big\{ \pmb{\phi}^H \pmb{v}^* \big\}.
\end{IEEEeqnarray}
\end{proposition}

\textbf{\textit{Proof: }} See \cite{pan2019intelligent,song2015sequence}. \hspace{4.5cm} $\blacksquare$

\begin{algorithm}[t!]\small
\caption{Joint Optimization of $\pmb{\ell}$, $\pmb{f}^e$, $\pmb{W}$ and $\pmb{\theta}$}
\begin{algorithmic}
 \renewcommand{\algorithmicrequire}{\textbf{Input:}}
 \renewcommand{\algorithmicensure}{\textbf{Output:}}
 \REQUIRE $\pmb{h}_{r,k}$, $\pmb{h}_{d,k}$, $\pmb{G}$, $B$, $p_t$, $\sigma^2$, $\varpi_k$, $L_k$, $c_k$, $f^e_{\text{total}}$, and $\epsilon$
 \ENSURE  Optimal $\pmb{\ell}$, $\pmb{f}^e$, $\pmb{W}$ and $\pmb{\theta}$
\\ \textbf{1. Initialization}
\STATE initialize $t_4 = 0$, $\epsilon_4^{(0)} = 1$
\STATE initialize $\pmb{\theta}^{(0)}$ satisfying \eqref{eqn:P0_constraint1} and ${\pmb{f}^e}^{(0)}$ satisfying \eqref{eqn:P0_constraint3} and \eqref{eqn:P0_constraint4}
\STATE calculate $\pmb{W}^{(0)}$ using \eqref{eqn:MMSE_MUD}
\\ \textbf{2. Joint optimization of $\pmb{\ell}$ and $\pmb{f}^e$, given $\pmb{W}^{(t_4)}$ and $\pmb{\theta}^{(t_4)}$}
\STATE calculate $\pmb{\ell}^{(t_4+1)}$ and ${\pmb{f}^e}^{(t_4+1)}$ using Algorithm~\ref{alg:joint_optimization1}
\\ \textbf{3. Joint optimization of $\pmb{W}$ and $\pmb{\theta}$, given $\pmb{\ell}^{(t_4+1)}$ and ${\pmb{f}^e}^{(t_4+1)}$}
\STATE calculate $\pmb{W}^{(t_4+1)}$ and ${\pmb{\theta}}^{(t_4+1)}$ using Algorithm~2
\\ \textbf{4. Convergence checking}
\STATE $\epsilon_4^{(t_4)} =\frac{ \big|\text{obj}\big(\pmb{\ell}^{(t_4+1)},{\pmb{f}^e}^{(t_4+1)},\pmb{W}^{(t_4+1)},{\pmb{\theta}}^{(t_4+1)}\big)- \text{obj}\big(\pmb{\ell}^{(t_4)},{\pmb{f}^e}^{(t_4)},\pmb{W}^{(t_4)},{\pmb{\theta}}^{(t_4)}\big)\big|}{\text{obj}\big(\pmb{\ell}^{(t_4+1)},{\pmb{f}^e}^{(t_4+1)},\pmb{W}^{(t_4+1)},{\pmb{\theta}}^{(t_4+1)}\big)}$
\IF{$\epsilon_4^{(t_4)} > \epsilon$ $\&\&$ $t_4 < t^{\text{max}}_4$ holds}
\STATE $t_4 = t_4+1$
\STATE Go to Step 2
\ELSE
\STATE integerize $\ell^{(t_4+1)}$ by \eqref{eqn:offloading_volume_calculation}
\STATE Output the optimal $\pmb{\ell}^*$, ${\pmb{f}^e}^*$, $\pmb{W}^*$ and $\pmb{\theta}^*$
\ENDIF
\end{algorithmic}
\label{alg:joint_optimization4}
\end{algorithm}

Here, the terms on the right side of \eqref{eqn:surrogate_function} is defined by our surrogate function $g(\pmb{\phi}|\pmb{\phi}^t)$. Then, Problem $\mathcal{P}2\text{-}E7$ at the $t$-th iteration is reformulated as
\begin{align}
\mathcal{P}2\text{-}E8: & \mathop {\min} \limits_{\pmb{\phi}} g(\pmb{\phi}|\pmb{\phi}^t) \nonumber \\
& \text{s.t.}  \quad  |\phi_n | = 1, \quad \forall n \in \{1,2,\ldots,N \}.
\end{align}
Since $(\pmb{\phi}^t)^H (\hat{\lambda}_{max} \pmb{I}_N - \pmb{\Psi}) \pmb{\phi}^t$ is a constant for a given $\pmb{\phi}^t$ and we have $\pmb{\phi}^H \hat{\lambda}_{max} \pmb{I}_N \pmb{\phi} = M \hat{\lambda}_{max}$, Problem $\mathcal{P}2\text{-}E8$ can be equivalently written as
\begin{align}
\mathcal{P}2\text{-}E9: & \mathop {\max} \limits_{\pmb{\phi}} \Re \Big\{ \pmb{\phi}^H \big[ (\hat{\lambda}_{max} \pmb{I}_N - \pmb{\Psi}) \pmb{\phi}^t  - \pmb{v}^* \big]  \Big\}  \nonumber \\
& \text{s.t.}  \quad  |\phi_n | = 1, \quad \forall n \in \{1,2,\ldots,N \}.
\end{align}
Then, the optimal solution of Problem $\mathcal{P}2\text{-}E9$ is readily given by
\begin{IEEEeqnarray}{rCl}\label{eq:phi_MM}
\pmb{\phi}^{t+1} = e^{j \arg \{ (\hat{\lambda}_{max} \pmb{I}_N - \pmb{\Psi}) \pmb{\phi}^t  - \pmb{v}^* \}}.
\end{IEEEeqnarray}
Accordingly, the optimal solution to Problem $\mathcal{P}2\text{-}E6$ can be obtained as
\begin{IEEEeqnarray}{rCl}
\pmb{\theta}^{t+1} = \arg \{ (\hat{\lambda}_{max} \pmb{I}_N - \pmb{\Psi}) \pmb{\phi}^t  - \pmb{v}^* \}.
\end{IEEEeqnarray}
The termination condition of the MM algorithm is given by $\big| f(\pmb{\phi}^{t+1}) - f(\pmb{\phi}^{t}) \big|/ f(\pmb{\phi}^{t+1}) \leq \epsilon$ or $t \geq t^{\text{max}}_{\text{MM}}$.
The procedure of solving Problem~$\mathcal{P}2\text{-}E3$ is summarized in Algorithm~\ref{alg:joint_optimization2}.

The complexity of Algorithm~\ref{alg:joint_optimization2} is dominated by its Step 2. Specifically, the complexity of calculating $\pmb{W}^{(t_3+1)}$ by \eqref{eqn:MMSE_MUD} is on the order of $\mathcal{O} \big(\max \{KM^3, KMN^2\} \big)$; the complexity of calculating $\pmb{\Upsilon}^{(t_3+1)}$ by \eqref{eqn:auxiliary_variable} is on the order of $\mathcal{O}(K)$. With regard to the calculation of ${\pmb{\theta}}^{(t_3+1)}$ using the MM algorithm, the complexity of calculating the eigenvalue $\lambda_{max}$ of $\pmb{\Psi}$ is on the order of $\mathcal{O}(N^3)$, while for each iteration of the MM algorithm, the main complexity lies in the calculation of $\pmb{\phi}^{t+1}$ in \eqref{eq:phi_MM}, whose complexity is on the order of $\mathcal{O}(N^2)$. Hence the complexity of the MM algorithm is $\mathcal{O}(N^3 + t^{\text{max}}_{\text{MM}} N^2)$. Summing these three terms together, we obtain the total complexity of Algorithm~\ref{alg:joint_optimization2} as $\mathcal{O}\big( \max \{N^3 + t^{\text{max}}_{\text{MM}} N^2, KM^3, KMN^2\} \big)$. Finally, the complexity of Algorithm~2 is mainly dependent on updating $\pmb{W}^{(t_2+1)}$ and $\pmb{\theta}^{(t_2+1)}$ using Algorithm~\ref{alg:joint_optimization2}, because all other steps are given by explicit mathematical expressions.

\subsection{Overall Algorithm to Solve Problem $\mathcal{P}0$}

Based on the above discussions, we provide the detailed description of the BCD algorithm used for solving Problem $\mathcal{P}0$ in Algorithm~\ref{alg:joint_optimization4}. Note that a decreasing OF value of Problem $\mathcal{P}0$ is guaranteed in Step 2 and Step 3. Furthermore, the OF value has a lower bound due to the constraint on the total edge computing resources. Hence, Algorithm~\ref{alg:joint_optimization4} is guaranteed to converge.

The computational complexity of Algorithm~\ref{alg:joint_optimization4} is mainly dependent on its Step 2 and Step 3, whose complexities have been analyzed in the above subsections. Furthermore, the simulation results in Section~\ref{sec:Numerical Results} show that Algorithm~\ref{alg:joint_optimization4} converges rapidly, which demonstrates the low complexity of our algorithms.

\section{Specific Case Study: The Single-Device Scenario}
\label{sec:Special Case Study}

In order to fully characterize the IRS-aided MEC system, a special case is investigated in this section, where a single device is served by the MEC system.
The optimization problem of the single-device scenario becomes much simpler for the following reasons. Firstly, the edge computing resource allocation no longer has to be considered, because all the edge computing resources can be assigned to this single device. Secondly, the sum-of-ratios form in Problem $\mathcal{P}2\text{-}E1$ becomes a single-ratio form, which implies that the optimization problem is more tractable. Thirdly, the multi-user interference does not have to be considered, when the detection vector and the IRS phase shift coefficient vector are optimized. The joint optimization is detailed as follows.

Problem $\mathcal{P}0$ can be simplified for the single-device scenario as
\begin{subequations}
\begin{align}
\mathcal{P}3: & \mathop {\min} \limits_{\pmb{w},\pmb{\theta},\ell} D(\pmb{w},\pmb{\theta},\ell) \nonumber \\
& \text{s.t.} \quad 0 \leq \theta_n < 2\pi, \quad n = 1,2,\ldots,N, \label{eqn:P3_constraint1} \\
& \quad \quad \ell \in \{0,1,\ldots,L\}, \label{eqn:P3_constraint2}
\end{align}
\end{subequations}
where the OF $D(\pmb{w},\pmb{\theta},\ell)$ becomes
\begin{IEEEeqnarray}{rCl}\label{eqn:single_user_latency}
D(\pmb{w},\pmb{\theta},\ell)
& = & \max \bigg\{ \frac{(L-\ell) c}{f^l}, \frac{\ell}{R(\pmb{w},\pmb{\theta})} + \frac{\ell c}{f^e_{\text{total}}} \bigg\}.
\end{IEEEeqnarray}
As illustrated in Proposition~\ref{proposition:offloading_volume}, for a given set of $\pmb{w}$ and $\pmb{\theta}$, $D(\pmb{w},\pmb{\theta},\ell)$ achieves its minimum value when $\ell$ is selected to ensure $ \frac{(L-\ell) c}{f^l} =  \frac{\ell}{R(\pmb{w},\pmb{\theta})} + \frac{\ell c}{f^e_{\text{total}}}$. Therefore, the optimal value of the relaxation of $\ell$ is given by
\begin{IEEEeqnarray}{rCl}\label{eqn:single_user_ell_hat}
\hat{\ell}^* = \frac{L c R f^e_{\text{total}}}{f^e_{\text{total}} f^l + c R \big(f^e_{\text{total}} + f^l \big)}.
\end{IEEEeqnarray}
Then, Problem $\mathcal{P}3$ is reformulated as
\begin{subequations}
\begin{align}
\mathcal{P}3\text{-}E1: & \mathop {\min} \limits_{\pmb{w},\pmb{\theta}}  \frac{\ell}{R(\pmb{w},\pmb{\theta})} + \frac{\ell c}{f^e_{\text{total}}} \nonumber \\
& \text{s.t.} \quad 0 \leq \theta_n < 2\pi, \quad n = 1,2,\ldots,N,
\end{align}
\end{subequations}
which is equivalent to
\begin{subequations}
\begin{align}
\mathcal{P}3\text{-}E2: & \mathop {\max} \limits_{\pmb{w},\pmb{\theta}} R(\pmb{w},\pmb{\theta}) \nonumber \\
& \text{s.t.} \quad 0 \leq \theta_n < 2\pi, \quad n = 1,2,\ldots,N.
\end{align}
\end{subequations}
Substituting \eqref{eqn:SINR} and \eqref{eqt:throughput} into the OF of Problem $\mathcal{P}3\text{-}E2$, and taking several steps of mathematical manipulation, Problem $\mathcal{P}3\text{-}E2$ may be equivalently transformed into
\begin{subequations}
\begin{align}
\mathcal{P}3\text{-}E3: & \mathop {\max} \limits_{\pmb{w},\pmb{\theta}}  \frac{p_t \big|\pmb{w}^H \big( \pmb{h}_{d} + \pmb{G} \pmb{\Theta} \pmb{h}_{r} \big) \big|^2}{\sigma^2 |\pmb{w}^H|^2 } \nonumber \\
& \text{s.t.} \quad 0 \leq \theta_n < 2\pi, \quad n = 1,2,\ldots,N.
\end{align}
\end{subequations}
Again, the BCD technique is invoked for optimizing $\pmb{w}$ and $\pmb{\theta}$ in Problem $\mathcal{P}3\text{-}E3$.
Specifically, given a $\pmb{\theta}$, $\pmb{w}$ can be optimized following the well-known maximum ratio combining (MRC) criterion \cite{goldsmith2005wireless}, which is given by
\begin{IEEEeqnarray}{rCl}\label{eqn:single_user_w}
\pmb{w} = \sqrt{p_t}( \pmb{h}_{d} + \pmb{G} \pmb{\Theta} \pmb{h}_{r})/\sigma,
\end{IEEEeqnarray}
while for a given $\pmb{w}$, we have the following inequality for the OF of Problem $\mathcal{P}3\text{-}E3$,
\begin{IEEEeqnarray}{rCl}\label{eqn:single_user_theta_equality}
\frac{p_t \big|\pmb{w}^H \big( \pmb{h}_{d} + \pmb{G} \pmb{\Theta} \pmb{h}_{r} \big) \big|^2}{\sigma^2 |\pmb{w}^H|^2 } \leq \frac{p_t \big|\pmb{w}^H \pmb{h}_{d} \big|^2}{\sigma^2 |\pmb{w}^H|^2 } + \frac{p_t \big|\pmb{w}^H \pmb{G} \pmb{\Theta} \pmb{h}_{r} \big|^2}{\sigma^2 |\pmb{w}^H|^2 }. \nonumber \\
\end{IEEEeqnarray}
The equality in \eqref{eqn:single_user_theta_equality} holds only when the IRS phase shift coefficient obeys $\arg\{\pmb{w}^H \pmb{h}_{d}\} = \arg\{\pmb{w}^H\pmb{G} \pmb{\Theta} \pmb{h}_{r}\}$. Accordingly, the reflection phase shift vector $\pmb{\theta}$ may be readily obtained as
\begin{IEEEeqnarray}{rCl}\label{eqn:single_user_theta}
\pmb{\theta} = \arg\{\pmb{w}^H \pmb{h}_{d}\} - \arg\{\text{diag}\{\pmb{w}^H\pmb{G}\} \pmb{h}_{r}\}.
\end{IEEEeqnarray}
In Algorithm~\ref{alg:joint_optimization5}, we provide the overall algorithm that is used for solving our optimization problem for the single-device scenario.

The complexity of Algorithm~\ref{alg:joint_optimization5} is dominated by calculating ${\pmb{\theta}}^{(t_5+1)}$ and $\pmb{w}^{(t_5+1)}$ using \eqref{eqn:single_user_theta} and \eqref{eqn:single_user_w}, whose complexities are on the order of $\mathcal{O}\big(\max\{MN,N^2\}\big)$ and of $\mathcal{O}(MN^2)$, respectively. Hence the complexity of Algorithm~\ref{alg:joint_optimization5} is on the order of $\mathcal{O}(MN^2)$.

\begin{spacing}{0.7}
\begin{algorithm}[t]\small
\caption{Joint Optimization of $\ell$, $\pmb{w}$ and $\pmb{\theta}$ proposed for the single-user scenario}
\begin{algorithmic}
 \renewcommand{\algorithmicrequire}{\textbf{Input:}}
 \renewcommand{\algorithmicensure}{\textbf{Output:}}
 \REQUIRE $\pmb{h}_{r}$, $\pmb{h}_{d}$, $\pmb{G}$, $B$, $p_t$, $\sigma^2$, $L$, $c$, $f^e_{\text{total}}$, and $\epsilon$
 \ENSURE  Optimal $\ell$, $\pmb{w}$ and $\pmb{\theta}$
\\ \textbf{1. Initialization}
\STATE initialize $t_5 = 0$, $\epsilon_5^{(0)} = 1$
\STATE initialize $\pmb{\theta}^{(0)}$ satisfying \eqref{eqn:P0_constraint1}
\STATE calculate $\pmb{w}^{(0)}$ using \eqref{eqn:single_user_w}
\\ \textbf{2. Joint optimization of $\pmb{w}$ and $\pmb{\theta}$}
\REPEAT
\STATE $\bullet$ calculate ${\pmb{\theta}}^{(t_5+1)}$ and $\pmb{w}^{(t_5+1)}$ using \eqref{eqn:single_user_theta} and \eqref{eqn:single_user_w}, respectively
\STATE $\bullet$ $\epsilon_5^{(t_5)} =\frac{ \big|\text{obj}\big(\pmb{w}^{(t_5+1)},{\pmb{\theta}}^{(t_5+1)}\big)- \text{obj}\big(\pmb{w}^{(t_5)},{\pmb{\theta}}^{(t_5)}\big)\big|}{\text{obj}\big(\pmb{w}^{(t_5+1)},{\pmb{\theta}}^{(t_5+1)}\big)}$,
where the $\text{obj}$ refers to the OF of Problem $\mathcal{P}3\text{-}E3$
\STATE $\bullet$ $t_5 = t_5+1$
\UNTIL $\epsilon_5^{(t_5)} \leq \epsilon$ $||$ $t_5 > t^{\text{max}}_5$
\\ \textbf{3. Optimization of $\pmb{\ell}$}
\STATE calculate $\hat{\ell}^{(t_5+1)}$ using \eqref{eqn:single_user_ell_hat}
\STATE integerize $\ell^{(t_5+1)}$ by \eqref{eqn:offloading_volume_calculation}
\end{algorithmic}
\label{alg:joint_optimization5}
\end{algorithm}
\end{spacing}

\section{Numerical Results}
\label{sec:Numerical Results}

\begin{figure*}[t]
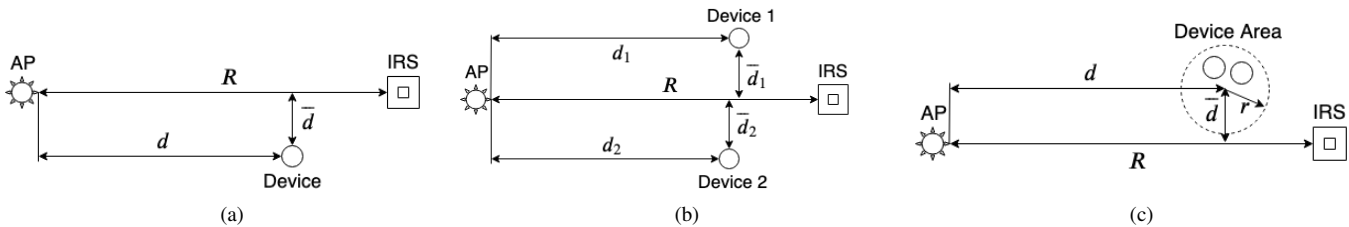
\center
\subfloat[]{\includegraphics[width=2.38in]{SU_diagram.png}\label{fig:su_diagram}}
\subfloat[]{\includegraphics[width=2.38in]{MU_diagram.png}\label{fig:mu_diagam}}
\subfloat[]{\includegraphics[width=2.38in]{MU_mu_diagram.png}\label{fig:mu_mu}}
\caption{Top-view setting of (a) the single-device scenario, (b) the two-device scenario, and (c) the multiple-device scenario.}
\label{fig:top_view}
\end{figure*}

In this section, the benefits of deploying the IRS in a MEC system are evaluated, relying on our algorithms developed in Section~\ref{sec:Solution to Problem} and \ref{sec:Special Case Study}. We consider a single-cell MEC system for both the single-device and two-device as well as multi-device scenarios. As shown in Fig.~\ref{fig:top_view}, the AP's coverage radius is $R=300~\rm{m}$ and the IRS is deployed at the cell edge. The location of the device is specified both by $d$ and by $\overline{d}$ in the single-device scenario, while in the two-device scenario, the devices' locations are specified by $(d_1,\overline{d}_1 )$ and $(d_2,\overline{d}_2)$, respectively. Furthermore, in the multiple-device scenario, it is assumed that the devices are uniformly distributed within a circle, whose size and location are prescribed by its radius $r$, as well as $d$ and $\overline{d}$, respectively. The default value of these parameters are set in the ``Location model" block of Table~\ref{tbl:simulation_parameters}.
As for the communications channel, we consider both the small scale fading and the large scale path loss. Specifically, the small scale fading is i.i.d. and obeys the complex Gaussian distribution associated with zero mean and unit variance, while the path loss in $\rm{dB}$ is given by
\begin{equation}
\text{PL} = \text{PL}_0 - 10 \alpha \log_{10} \Big( \frac{d}{d_0} \Big),
\end{equation}
where $\text{PL}_0$ is the path loss at the reference distance $d_0$; $d$ and $\alpha$ represent the distance of the communications link and its path loss exponent, respectively. Here we use $\alpha_{ua}$, $\alpha_{ui}$ and $\alpha_{ia}$ to denote the path loss exponent of the link between the device and the AP, that of the link between the device and the IRS, as well as that of the link between the IRS and the AP, respectively. The zero-mean additive white Gaussian noise associated with the variable of $\sigma^2$ is imposed on the off-loaded signal. The default settings of these parameters are specified in the ``Communications model" block of Table~\ref{tbl:simulation_parameters}.
The variables $L_k$, $c_k$ and $f^l_k$ obey the uniform distribution, whose ranges are given in the ``Computing model" block of Table~\ref{tbl:simulation_parameters}.

\begin{table}[h!]\small
\begin{center}
\caption{Default simulation parameter setting}
\label{tbl:simulation_parameters}
\begin{tabular}{ | l | r |}
\hline
Description & Parameter and Value \\ \hline \hline
\multirow{2}{*}{Location model}
& $R = 300~\rm{m}$ \\
& $\overline{d} = \overline{d}_1 = \overline{d}_2 = 10 ~\rm{m}$ \\ \hline
\multirow{6}{*}{Communication model}
& $\text{Bandwidth} = 1~\rm{MHz}$ \\
& $\text{PL}_0 = 30~\rm{dB}$, $d_0 = 1~\rm{m}$ \\
& $\alpha_{ua} = 3.5$, $\alpha_{ui} = 2.2$, $\alpha_{ia} = 2.2$  \\
& $M = 5$ \\
& $p_t = 1~\rm{mW}$ \\
& $\sigma^2 = 3.98\times 10^{-12}~\rm{mW}$ \\ \hline
\multirow{3}{*}{Computing model}
& $L_k = [250, 350]~\rm{Kb}$ \\
& $c_k = [700,800]~\rm{cycle/bit}$ \\
& $f^l_k = [4,6] \times 10^8~\rm{cycle/s}$ \\ \hline
Weight & $\varpi_k = 1/K$ \\ \hline
Convergence criterion &  $\epsilon = 0.001$ \\ \hline
\end{tabular}
\end{center}
\end{table}

\begin{figure}[t]\center
\subfloat[]{\includegraphics[width=3.6in]{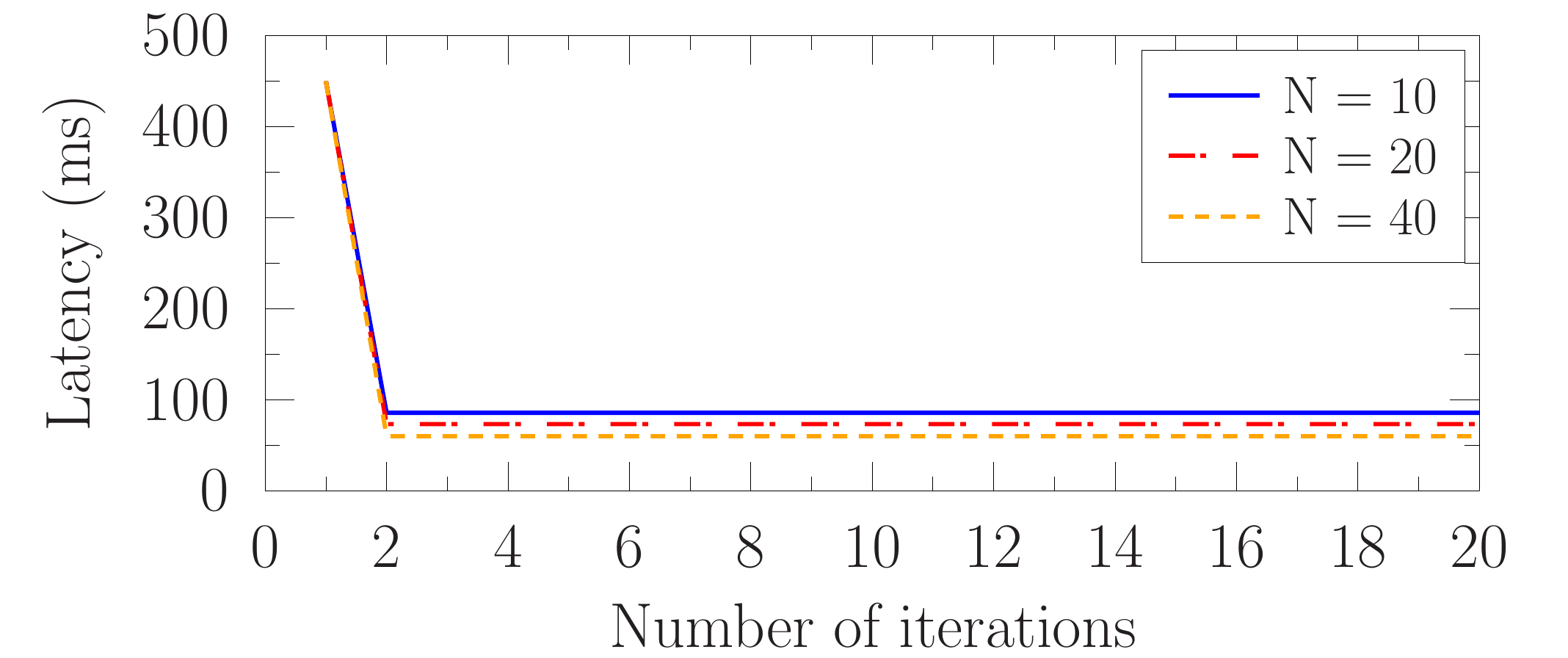}\label{fig:su_convergence}}\\
\subfloat[]{\includegraphics[width=3.6in]{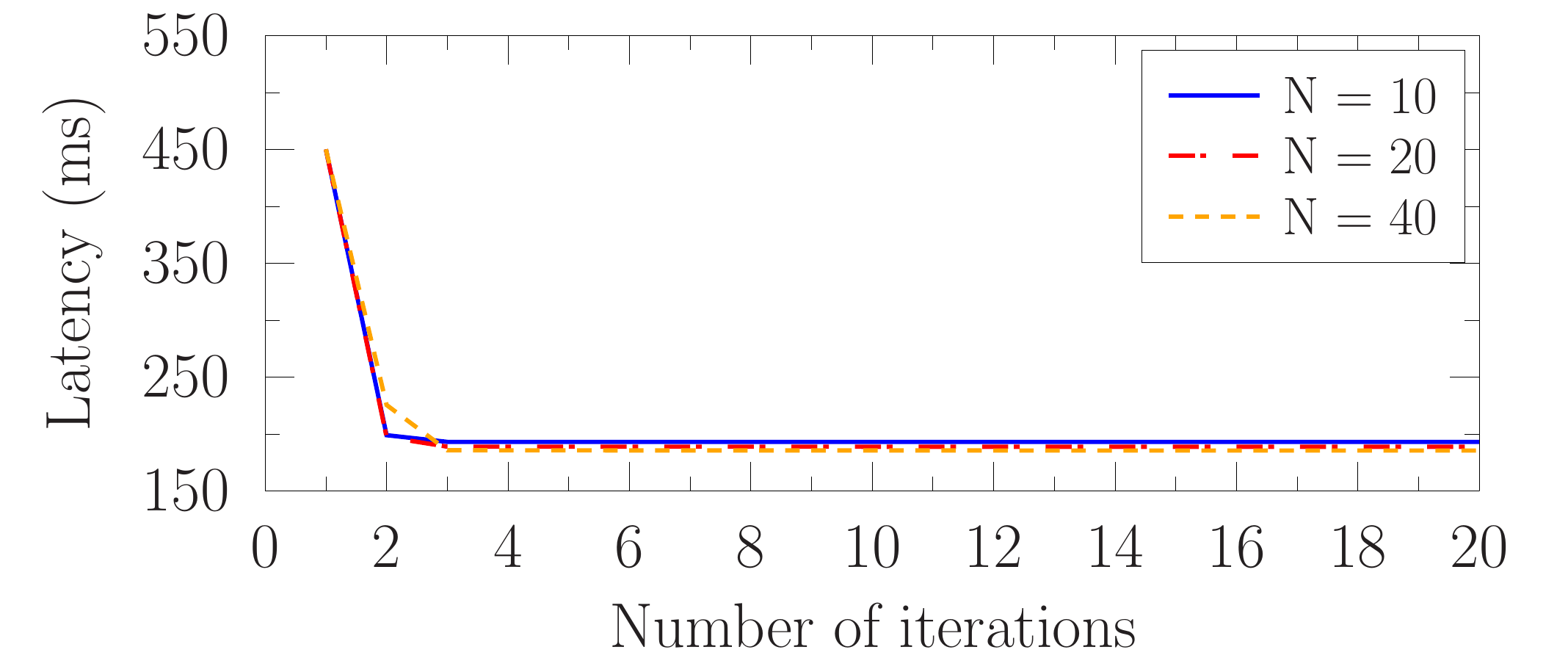}\label{fig:mu_convergence}}
\caption{Convergence of the algorithms for (a) the single-device scenario using Algorithm~\ref{alg:joint_optimization5} and (b) the two-device scenario using Algorithm~\ref{alg:joint_optimization4}. The parameters are set as follows: $f^e_{\text{total}} = 50\times 10^9~\rm{cycle/s}$; $c_k = 750~\text{cycle}$, $L_k = 300~\rm{Kb}$, and $f^l_k = 0.5\times 10^9~\rm{cycle/s}$ for all devices. (a): $d = 280~\rm{m}$; (b): $d_1 = d_2 = 280~\rm{m}$.}
\label{fig:convergence}
\end{figure}

The following subsections detail our simulation results, in terms of the properties of our proposed algorithm and of the latency performance in both the single-device, and two-device, as well as multi-device scenarios in various simulation environments.
The following three schemes are considered:
\begin{itemize}
\item \emph{With IRS:} The off-loading volume, edge computing resource allocation, MUD matrix, and IRS phase shift are optimized relying on Algorithm~\ref{alg:joint_optimization4} and Algorithm~\ref{alg:joint_optimization5} in the multi-device and single-device scenarios, respectively.
\item \emph{RandPhase:} The off-loading volume, edge computing resource allocation, as well as MUD matrix are optimized using Algorithm~\ref{alg:joint_optimization4} and Algorithm~\ref{alg:joint_optimization5} in multi-device and single-device scenarios, respectively, while skipping the step of designing the IRS phase shift, which is randomly set obeying the uniform distribution in the range of $[0,2\pi)$.
\item \emph{Without IRS:} The composite channel $\pmb{G}\pmb{\Theta}\pmb{h}_{r,k}$ taking into account the IRS is set to $0$. The off-loading volume, edge computing resource allocation, and the MUD matrix are designed following Algorithm~\ref{alg:joint_optimization4} and Algorithm~\ref{alg:joint_optimization5} in the multi-device and single-device scenarios, respectively.
\end{itemize}

\subsection{Properties of the Proposed Algorithms}

In this subsection, the properties of Algorithm~\ref{alg:joint_optimization4} and \ref{alg:joint_optimization5} are investigated, with the aid of numerical results.

\subsubsection{Convergence}

Fig.~\ref{fig:convergence} shows the device-average latency versus the number of iterations under various settings of the IRS phase shift number, i.e. $N=10$, $20$, and $40$, for both the single-device and multi-device scenarios. We have the following two observations. Firstly, a larger number of phase shifts leads to a slightly slower convergence, especially for the multi-device scenario. This is because more optimizing variables are involved. Secondly, the proposed algorithms are capable of achieving a convergence within $5$ iterations, which validates its practical implementation.

\subsubsection{Impact of the Initialization Settings}

\begin{figure}[t]\center
\subfloat[]{\includegraphics[width=3.6in]{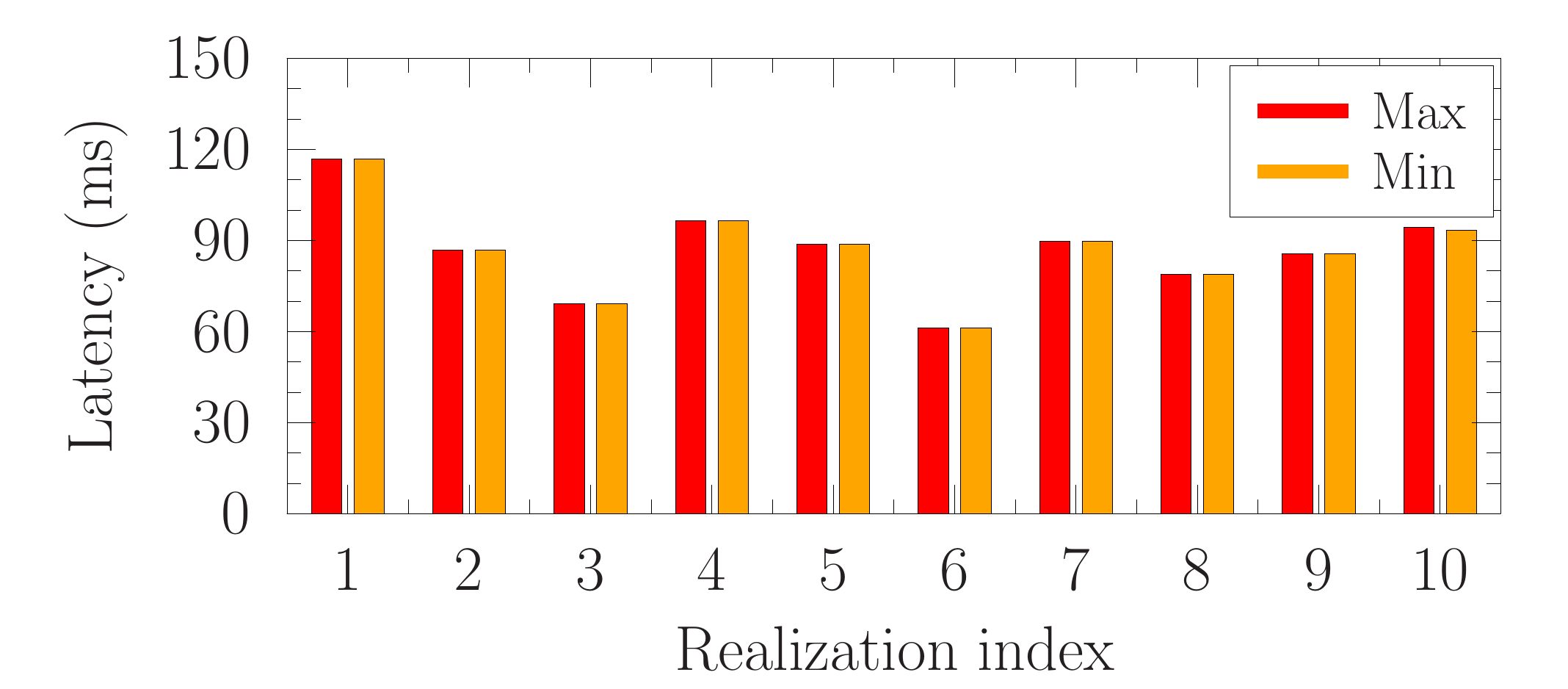}\label{fig:su_initialization}}\\
\subfloat[]{\includegraphics[width=3.6in]{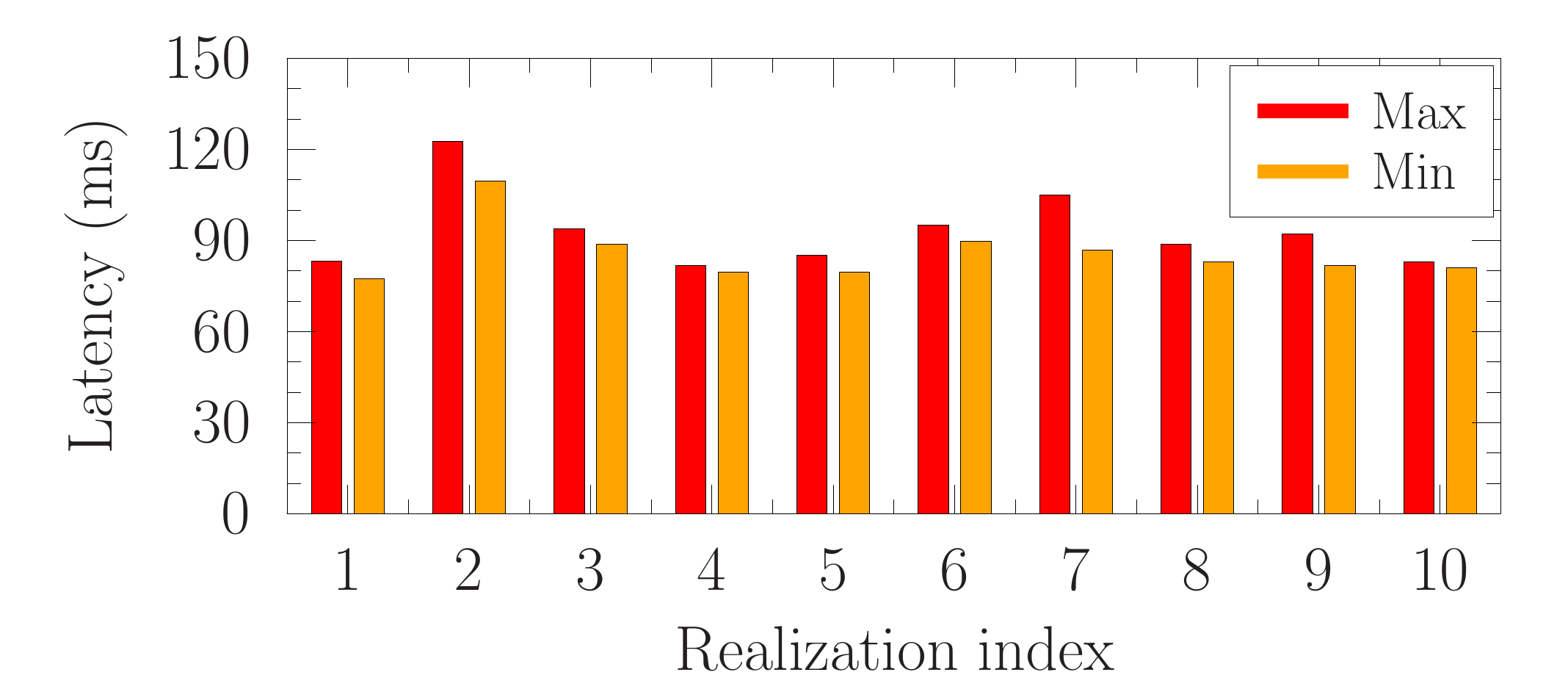}\label{fig:mu_initialization}}
\caption{Simulation results of the the maximum and minimum latency versus the realization index obtained under $100$ random initialization settings for (a) the single-device scenario using Algorithm~\ref{alg:joint_optimization5} and (b) the two-device scenario using Algorithm~\ref{alg:joint_optimization4}. ``Max" and ``Min" refer to the maximum and minimum value, respectively. The parameters are set as follows: $N = 40$. (a): $d = 280~\rm{m}$; (b): $d_1 = d_2 = 280~\rm{m}$.}
\label{fig:initialization}
\end{figure}
As elaborated on in Remark 1, locally optimal results are provided by our proposed algorithms. Hence, the results obtained are directly dependent on the initialization settings of Algorithm~\ref{alg:joint_optimization4} and \ref{alg:joint_optimization5}. In order to clarify its impact, Fig.~\ref{fig:initialization} presents the latency performance under different initialization settings both for single- and multi-device scenarios.
Specifically, for each realization of the wireless channels and computing tasks to be processed, $100$ locally optimal results are obtained using our proposed algorithms, where each of the initializations is randomly set. Among these locally optimal results, the maximum latency value that can be deemed to be the worst-case result using our proposed algorithms is labeled as ``Max" in Fig.~\ref{fig:initialization}, while the minimum latency value is labeled by ``Min" in Fig.~\ref{fig:initialization} which may resemble the globally optimal result. It is shown that these two values are almost identical for the single-device scenario, while their gap ranges from $2\%$ to $17\%$ in the multi-device scenario, which implies that our proposed algorithms are capable of approaching the optimal performance.

\subsubsection{Impact of the Phase Quantization}

\begin{figure}[t]\center
\subfloat[]{\includegraphics[width=3.6in]{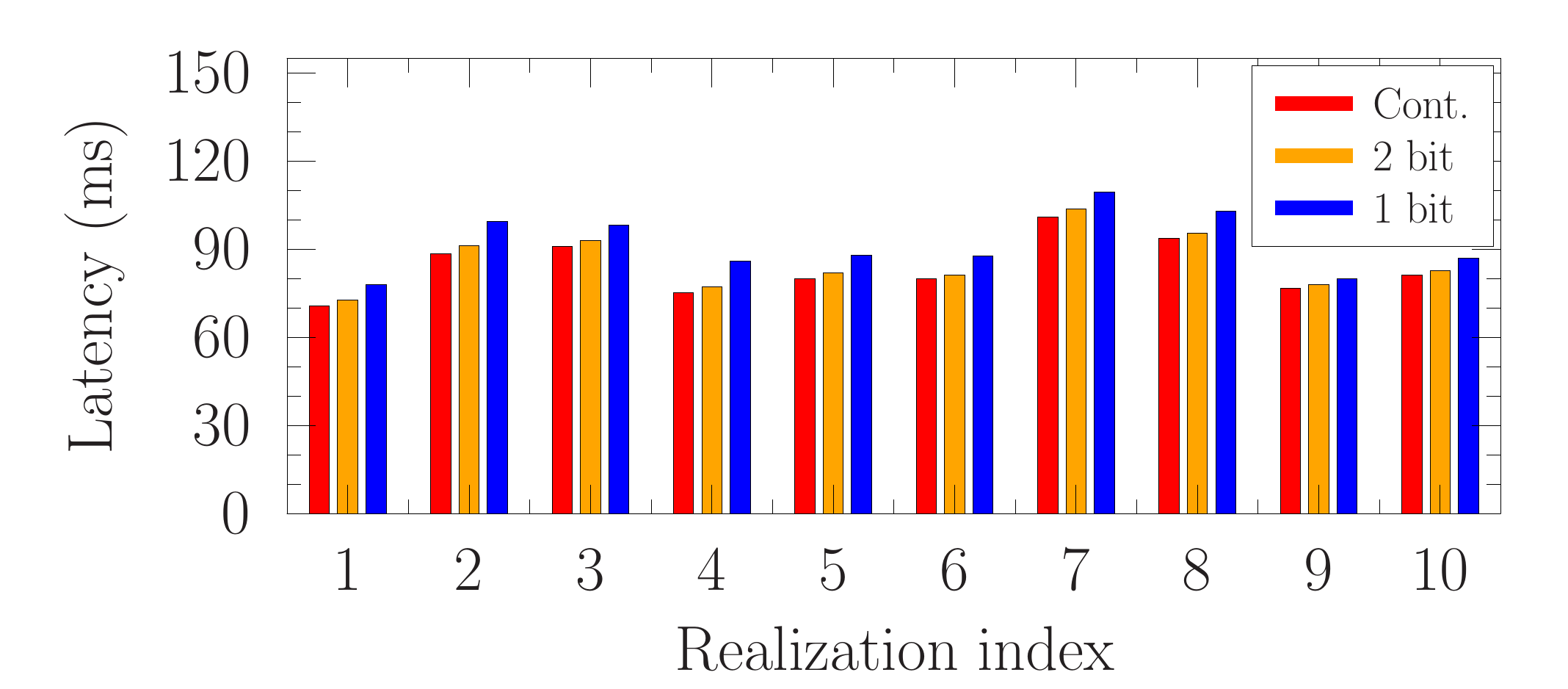}\label{fig:su_quantization}}\\
\subfloat[]{\includegraphics[width=3.6in]{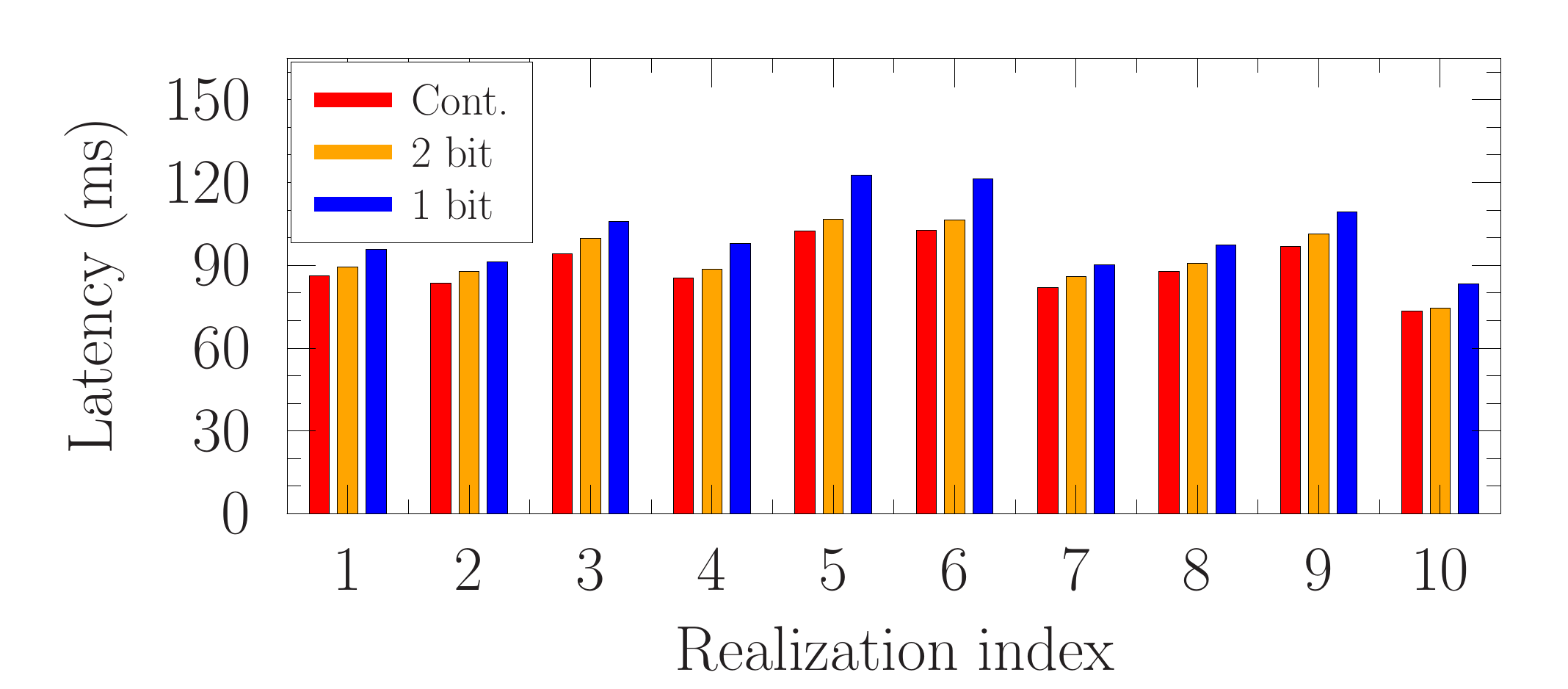}\label{fig:mu_quantization}}
\caption{Simulation results of the latency versus the realization index under different assumptions of IRS phase shifts for (a) the single-device scenario and (b) the two-device scenario. ``Cont.", ``1-bit" and ``2-bit" refer to the assumptions of continuous, 1-bit, and 2-bit phase shifts, respectively. The parameters are set as follows: $N = 40$. (a): $d = 280~\rm{m}$; (b): $d_1 = d_2 = 280~\rm{m}$.}
\label{fig:quantization}
\end{figure}

Due to the associated hardware limitation, only a limited number of discrete IRS phase shifts can be provided in practice \cite{wu2019beamforming}, which prohibits the direct implementation of our proposed algorithms. An intuitive practical solution to this issue is to round the continuous phase shift obtained to its nearest discrete phase shift. Naturally, a performance loss is imposed, owing to the associated quantization effect. Fig.~\ref{fig:quantization} evaluates the impact of phase quantization on the latency, where three practical assumptions are considered. Specifically, under the assumption of continuous phase shifts, the phase shift of each IRS element can be set as an arbitrary value in the interval of $[0,2\pi]$; Determined by a 1-bit control signal, the phase shift of each IRS element has to be either $0$ or $\pi$ under the assumption of 1-bit phase shift; for a 2-bit control signal, the phase shift of each IRS element has to be one of the values in the set of $\left \{ 0,\frac{\pi}{2},\pi,\frac{3\pi}{2}\right\}$. Particular to the schemes under discrete phase shift assumptions, the values of $\pmb{W}$, $\pmb{\ell}$, and $\pmb{f}^e$ are updated based on the quantized phase shifts and then the latency is calculated accordingly. We have the following observations. Firstly, as expected, the latency decreases upon increasing the number of discrete phase shifts. Secondly, the performance gap between the schemes under the assumptions of continuous phase shifts and $2$-bit phase shifts ranges from $1\%$ to $5\%$, which implies that the quantization loss becomes negligible for as few as four phase shifts in practice.

\subsection{Single-Device Scenario}

\begin{figure}[t!]\center
\includegraphics[width=0.47\textwidth]{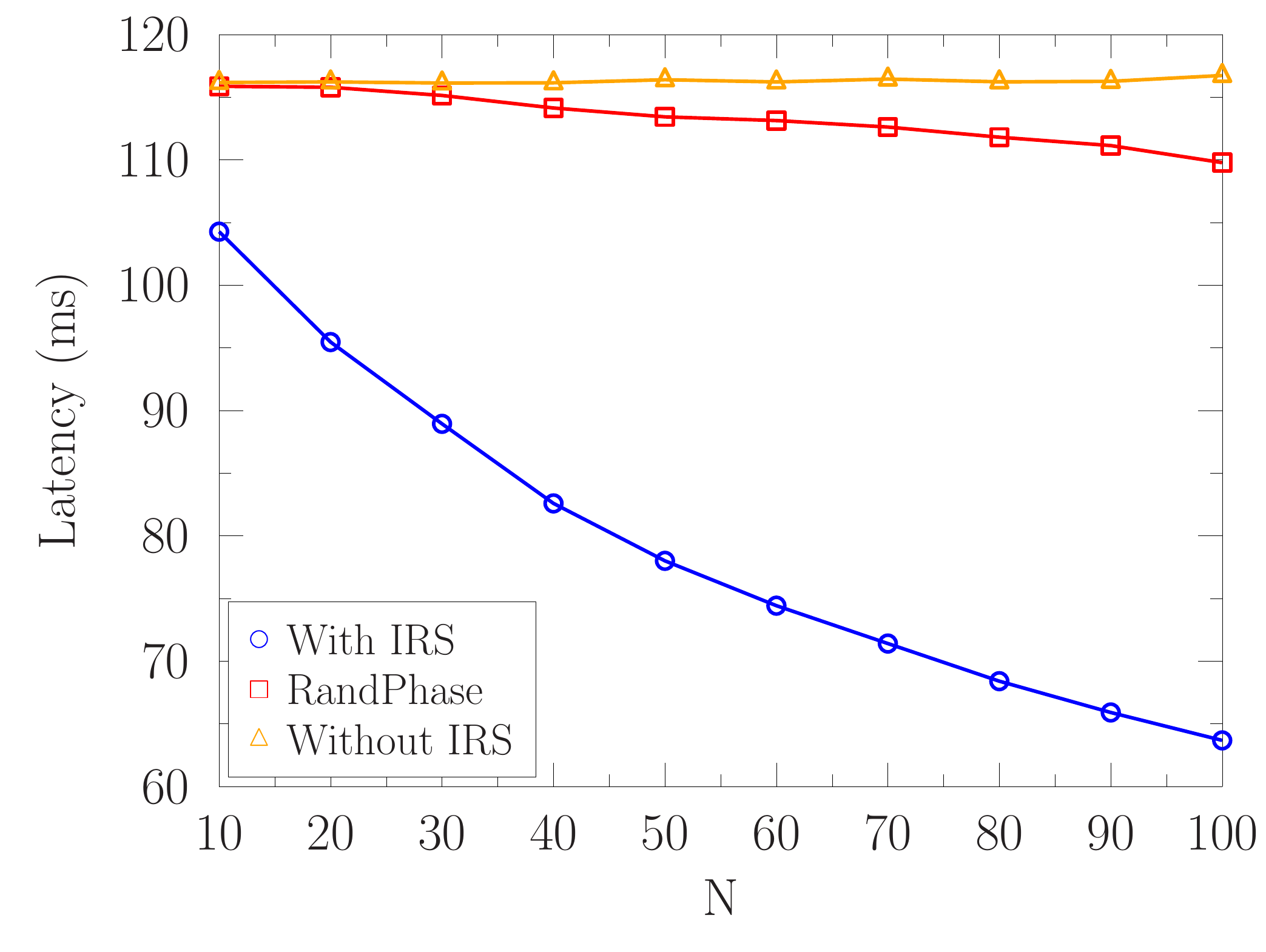}
\caption{Simulation results of the latency versus the number of the IRS elements in the single-device scenario, where we set $d = 280~\rm{m}$ and $f^e_{\text{total}} = 50\times 10^9~\rm{cycle/s}$.}
\label{fig:su_N}
\end{figure}

\begin{figure}[t!]\center
\includegraphics[width=0.47\textwidth]{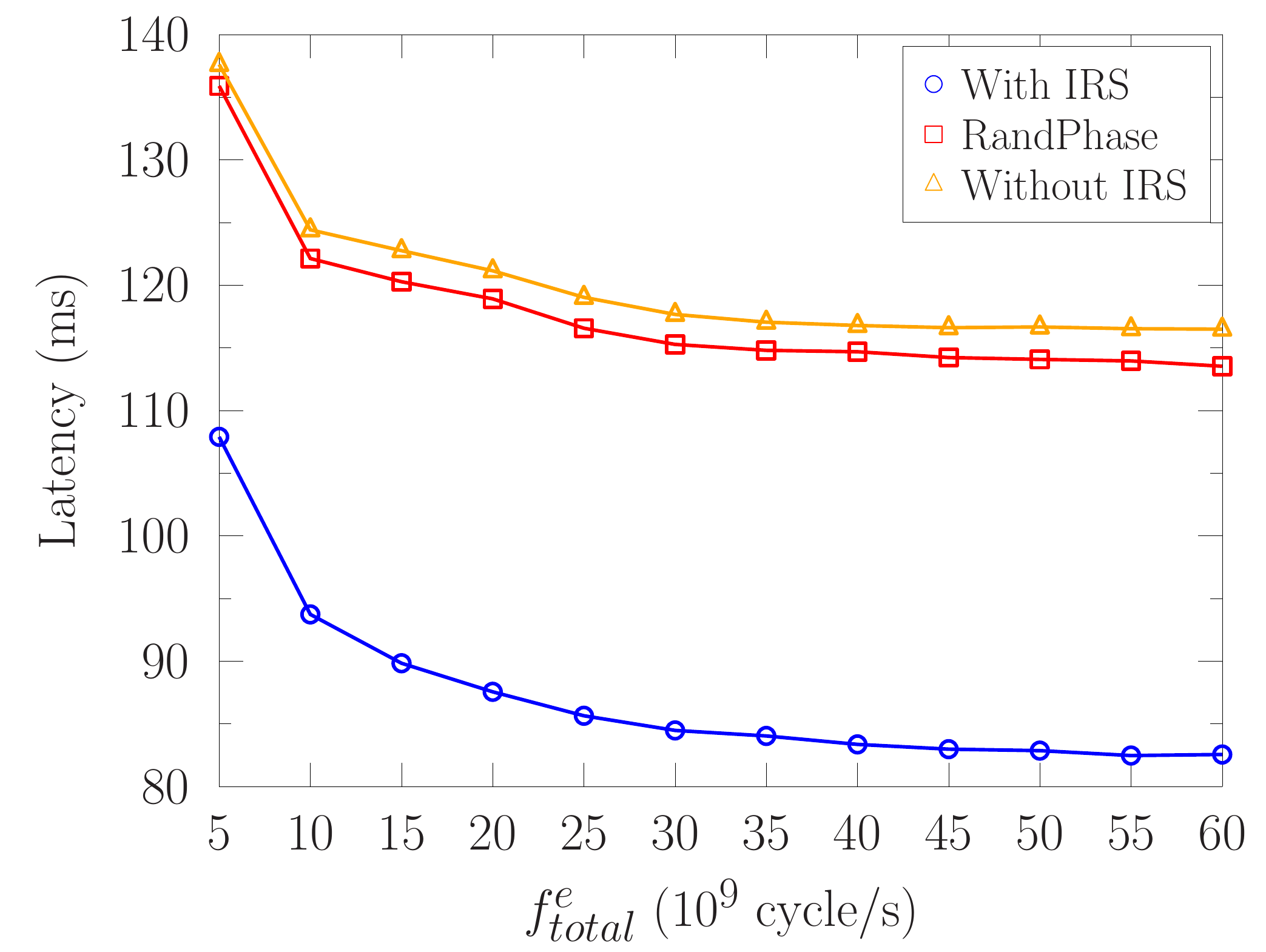}
\caption{Simulation results of the latency versus the edge computing capability in the single-device scenario, where we set $d = 280~\rm{m}$ and $N = 40$.}
\label{fig:su_f_total}
\end{figure}

Fig.~\ref{fig:su_N}-\ref{fig:su_distance} present the latency versus various parameter settings in the single-device scenario, discussed as follows.

\subsubsection{Impact of the Number of Reflecting Elements}
Fig.~\ref{fig:su_N} presents the latency versus the number of the reflecting elements, for the various phase shift design schemes. Our observations are as follows. Firstly, the performance gap between the schemes ``Without IRS" and ``RandPhase" becomes higher upon increasing the number of reflecting elements, which implies that the IRS is capable of assisting the computation off-loading even without carefully designing the phase shift. This is because the received SINR can be improved by deploying an IRS for computation off-loading. The gain was termed as the virtual array gain in Section~\ref{sec:Introduction}. Secondly, the performance gain of the scheme ``With IRS" over the scheme ``RandPhase" is around $11~\rm{ms}$ when we set $N = 10$, while it becomes $46~\rm{ms}$ when we have $N = 100$. This implies that a sophisticated design of the IRS phase shift response provides a beamforming gain, and that increasing the number of IRS elements leads to a higher reflection-based beamforming gain. Combining these two types of gains together, IRSs are capable of efficiently reducing the latency in MEC systems.

\subsubsection{Impact of the Edge Computing Capability}
Fig.~\ref{fig:su_f_total} shows the latency versus the edge computing capability, for various IRS phase shift schemes. Our observations are as follows. For all these three schemes, the increase of $f^e_{\text{total}}$ drastically reduces the latency when $f^e_{\text{total}}$ is of a small value, while the reduction of the latency becomes smaller when $f^e_{\text{total}}$ reaches a certain threshold value, say $30\times 10^9~\rm{cycle/s}$. This is because the latency imposed by the edge computing dominates when $f^e_{\text{total}}$ is of a small value, whereas the latency imposed by computation off-loading plays a dominant role when $f^e_{\text{total}}$ reaches a high value. Therefore, it is not necessary to equip the edge computing node with an extremely powerful computing capability for latency minimization.

\begin{figure}[t!]\center
\includegraphics[width=0.47\textwidth]{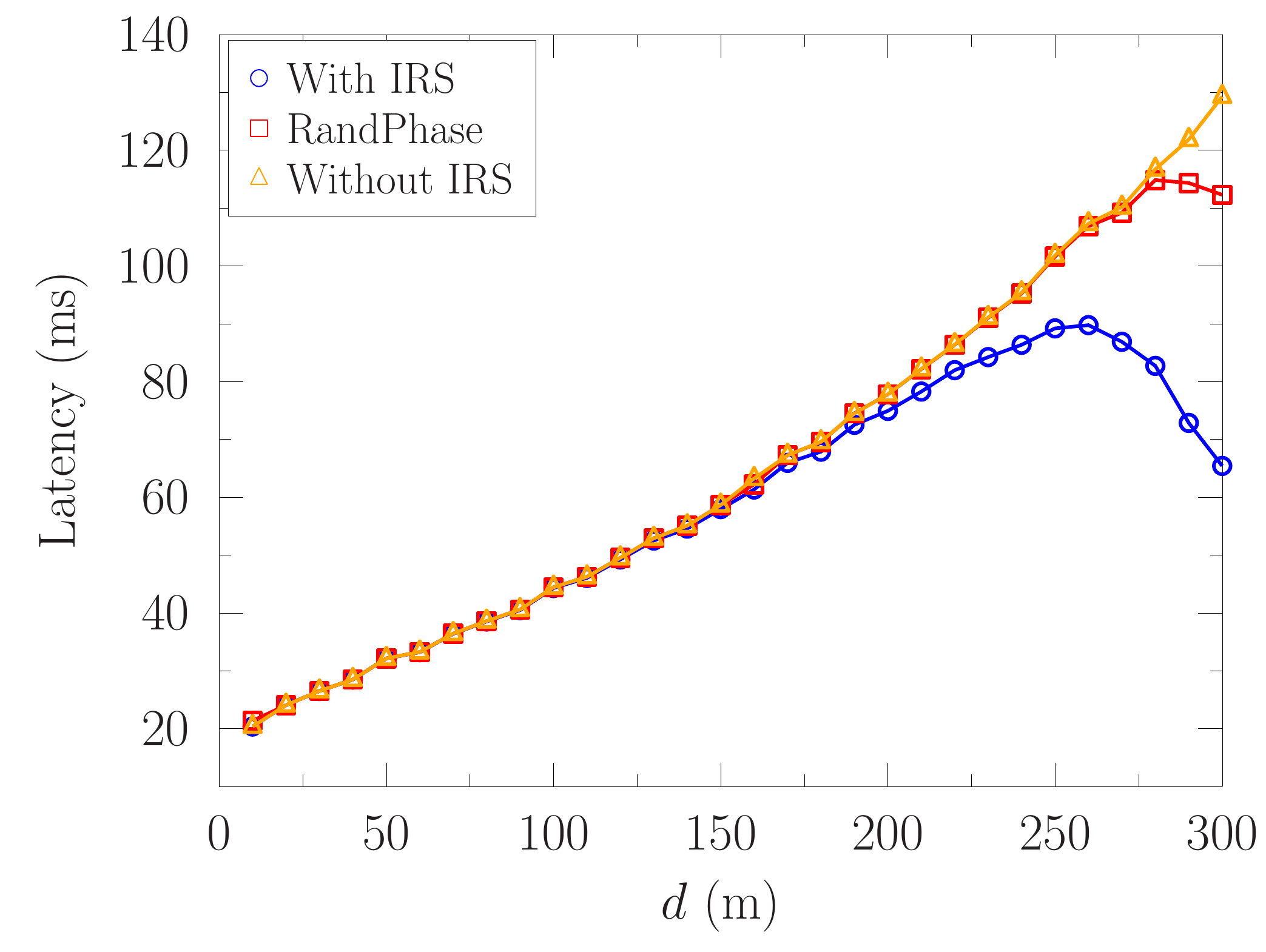}
\caption{Simulation results of the latency versus the device location in the single-device scenario, where we set $N = 40$ and $f^e_{\text{total}} = 50\times 10^9~\rm{cycle/s}$.}
\label{fig:su_distance}
\end{figure}

\subsubsection{Impact of the Device Location}
Fig.~\ref{fig:su_distance} depicts the latency versus the device location, equipped with various IRS phase shift schemes. Our observations are as follows. In the case where no IRS is employed, the latency increases upon increasing the distance between the AP and the device.
In the case where the IRS's phase shift is randomly set, the advantage of using the IRS becomes visible when the distance between the device and the IRS is less than $20~\rm{m}$. By contrast, the benefit of the IRS becomes notable for a much larger coverage of $100~\rm{m}$ for the ``With IRS" scheme. This observation implies that a sophisticated design of the IRS phase shift response is capable of extending the coverage of the IRS.
Furthermore, the latency reaches its maximum value at $d=260~\rm{m}$ and thereafter becomes smaller for the ``With IRS" scheme. This is because the direct device-AP link dominates the computation off-loading when the device's location obeys $d\leq 260~\rm{m}$, while the composite device-IRS-AP link plays a dominant role, when we have $d\geq 260~\rm{m}$. This observation further consolidates that a higher gain can be achieved in the near-IRS area, where the composite device-IRS-AP link dominates the computation off-loading.

\subsection{Multi-Device Scenario}

\begin{figure}[t!]\center
\includegraphics[width=0.47\textwidth]{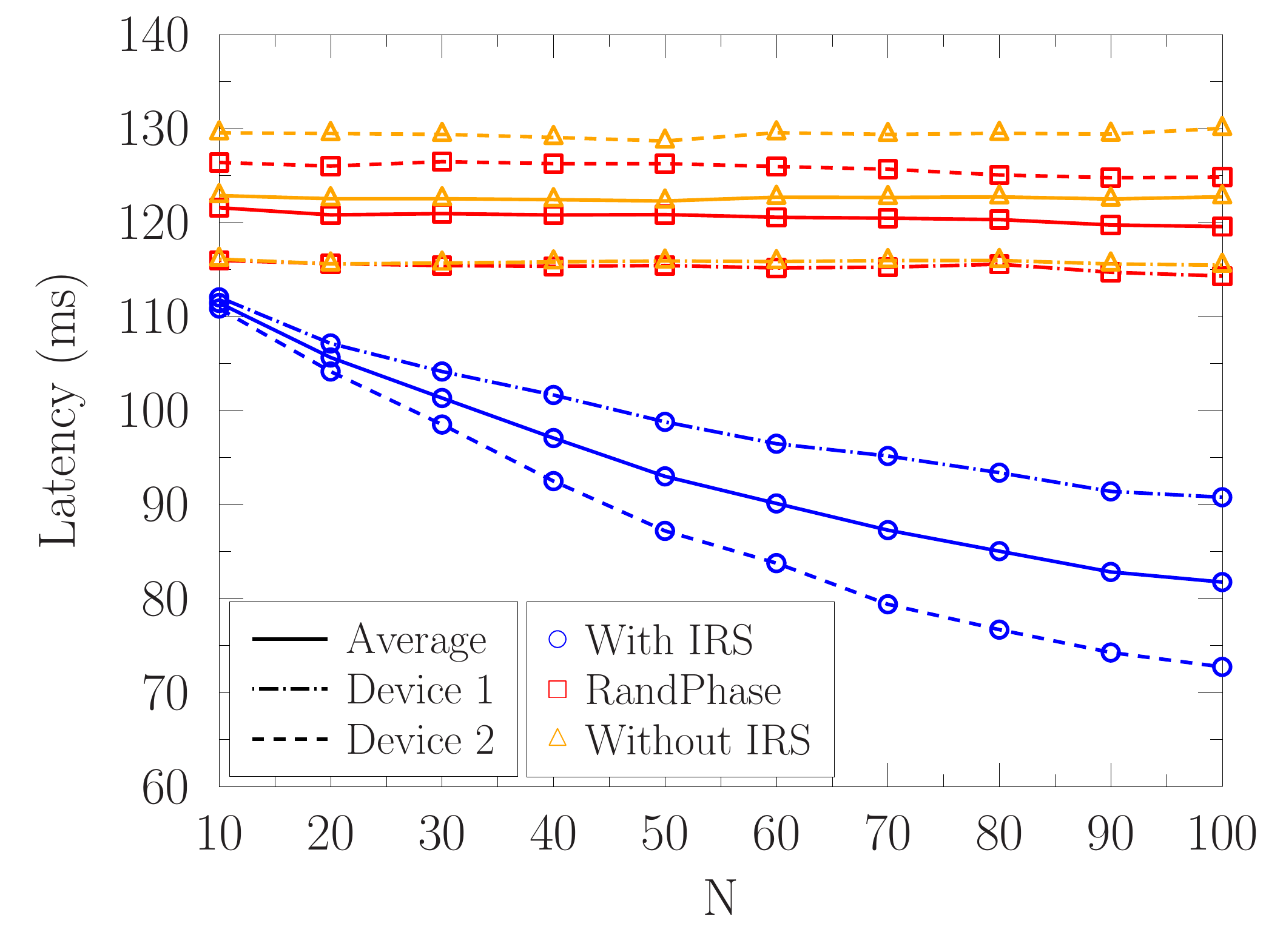}
\caption{Simulation results of the latency versus the number of the IRS elements in the two-device scenario, where we set $d_1 = 260~\rm{m}$, $d_2 = 280~\rm{m}$, and $f^e_{\text{total}} = 50\times 10^9~\rm{cycle/s}$.}
\label{fig:mu_N}
\end{figure}

\begin{figure}[t!]\center
\includegraphics[width=0.47\textwidth]{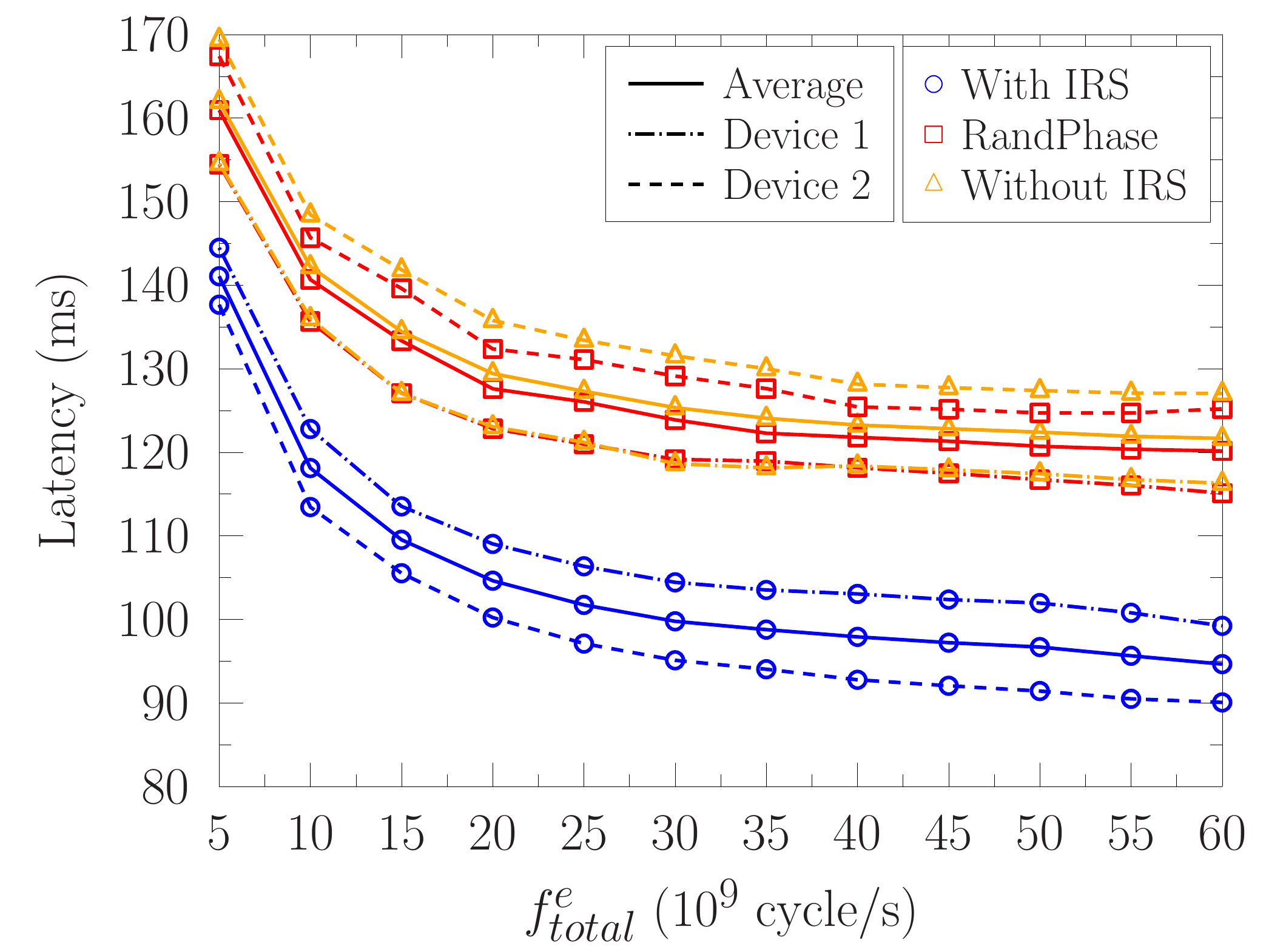}
\caption{Simulation results of the latency versus the edge computing resource in the two-device scenario, where we set $d_1 = 260~\rm{m}$, $d_2 = 280~\rm{m}$, and $N = 40$.}
\label{fig:mu_f_total}
\end{figure}

Fig.~\ref{fig:mu_N}-\ref{fig:mu_alpha} present the latency in the two-device scenario, and Fig.~\ref{fig:mu_mu}-\ref{fig:mu_ICI} show the latency in the multiple-device scenario, which are discussed as follows.

\subsubsection{Impact of the Number of IRS Elements}

Fig.~\ref{fig:mu_N} depicts the latency versus the number of the IRS elements in the two-device scenario, equipped with various phase shift schemes. Apart from the insights obtained in the single-device scenario, we also have the following observations. Firstly, Device 2 outperforms Device 1 for the ``With IRS" scheme, whilst Device 1 has a lower latency both for the ``Without IRS" and  ``RandPhase" schemes compared to Device 2. This is in accordance with the comparative relationship between the devices located at $d=260~\rm{m}$ and at $d=280~\rm{m}$ in terms of the latency using those three phase shift schemes, as shown in Fig.~\ref{fig:su_distance}. This also implies that IRSs may change the latency ranking of the devices in MEC systems. Secondly, upon increasing the number of IRS elements, Device 2 obtains a higher gain than Device 1. This is because Device 2 is located closer to the IRS, where the composite device-IRS-AP channel dominates the computation off-loading. Again, this implies that given a specific path loss exponent, a higher array and passive beamforming gain may be achieved if the device is located closer to the IRS.

\subsubsection{Impact of the Edge Computing Capability}

Fig.~\ref{fig:mu_f_total} presents the latency versus the edge computing capability in the multi-device scenario, equipped with various phase shift schemes. This scenario follows similar trends to the single-device case illustrated in Fig.~\ref{fig:su_f_total}.

\begin{figure}[t!]\center
\includegraphics[width=0.47\textwidth]{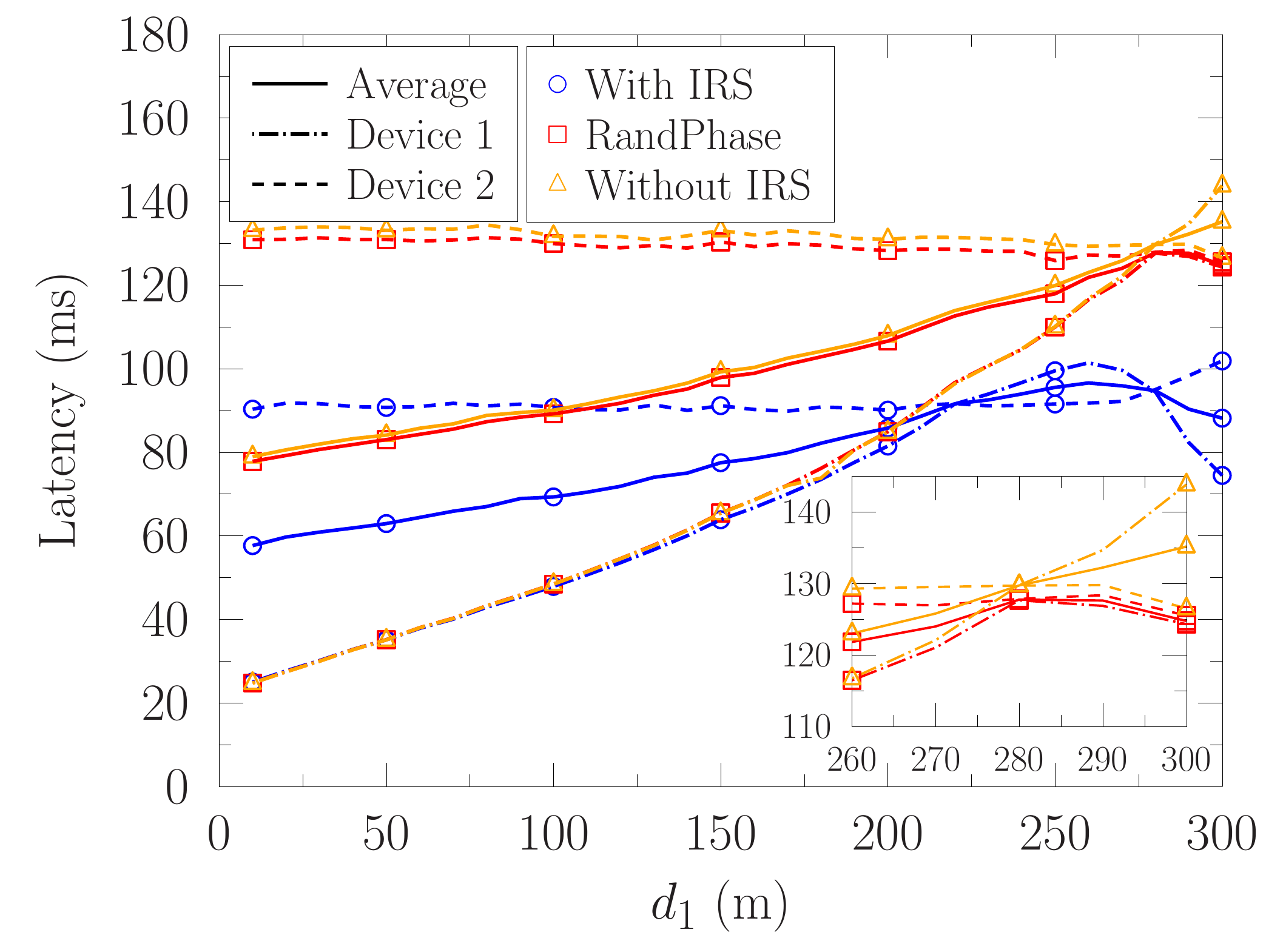}
\caption{Simulation results of the latency versus the user location in the two-device scenario, where we set $d_2 = 280~\rm{m}$, $N = 40$, and $f^e_{\text{total}} = 50\times 10^9~\rm{cycle/s}$.}
\label{fig:mu_distance}
\end{figure}

\subsubsection{Impact of the Device Location}

Fig.~\ref{fig:mu_distance} plots the latency versus the location of Device 1, while fixing the location of the AP, the IRS, and Device 2.
As for the ``With IRS" scheme, the curve of Device 1's latency intercepts that of Device 2 at $d_1=220~\rm{m}$ and $d_1=280~\rm{m}$, which implies that the devices at these two locations have the same channel gain. In other words, the IRS is capable of assisting the device at $d=280~\rm{m}$ to achieve the same latency as the device at $d=220~\rm{m}$. Note that the specific values of these two equivalent-latency locations are dependent on the specific values of the path loss exponents of the device-IRS, IRS-AP, and device-AP channels, as presented below.

\begin{figure}[t!]\center
  \includegraphics[width=0.47\textwidth]{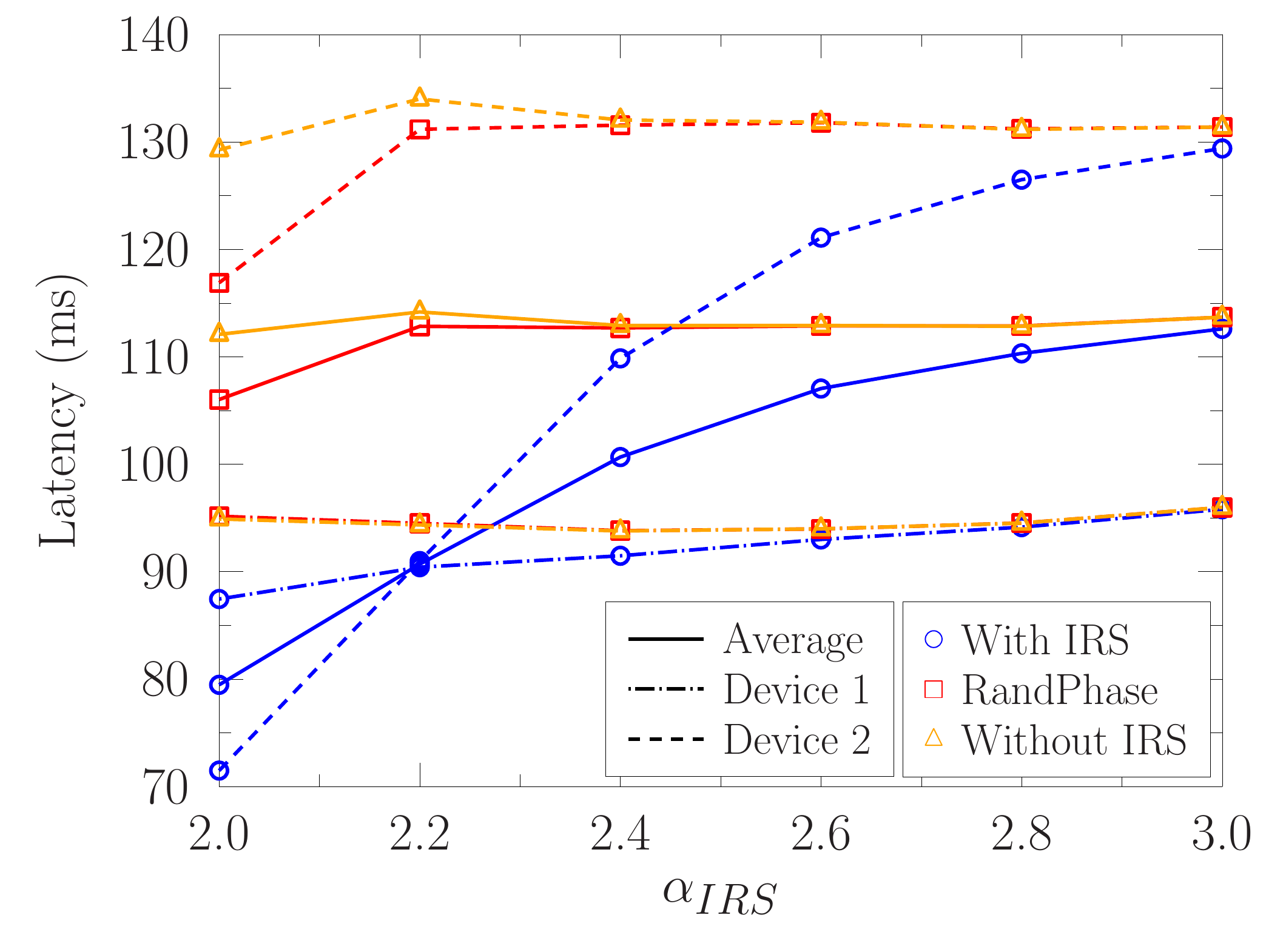}
  \captionof{figure}{Simulation results of the latency versus the path loss exponent in the two-device scenario, where we set $\alpha_{ui} = \alpha_{ia} = \alpha_{\text{IRS}}$. The parameters are set as follows: $d_1 = 220~\rm{m}$, $d_2 = 280~\rm{m}$, $N = 40$, and $f^e_{\text{total}} = 50\times 10^9~\rm{cycle/s}$.}
  \label{fig:mu_alpha}
\end{figure}

\subsubsection{Impact of the Path Loss Exponent}

Fig.~\ref{fig:mu_alpha} illustrates the latency versus the path loss exponent value associated with the IRS. It can be observed that the intercept point disappears, when $\alpha_{\text{IRS}}$ is changed for the ``With IRS" scheme. Furthermore, the latency of devices increases upon increasing $\alpha_{\text{IRS}}$. This is because higher $\alpha_{\text{IRS}}$ leads to a lower array and beamforming gain by the IRS. This provides important insights for engineering design: the location of the IRS should be carefully selected to avoid obstacles, for achieving a lower $\alpha_{ui}$ and $\alpha_{ia}$.

\begin{figure}[t!]\center
  \includegraphics[width=0.47\textwidth]{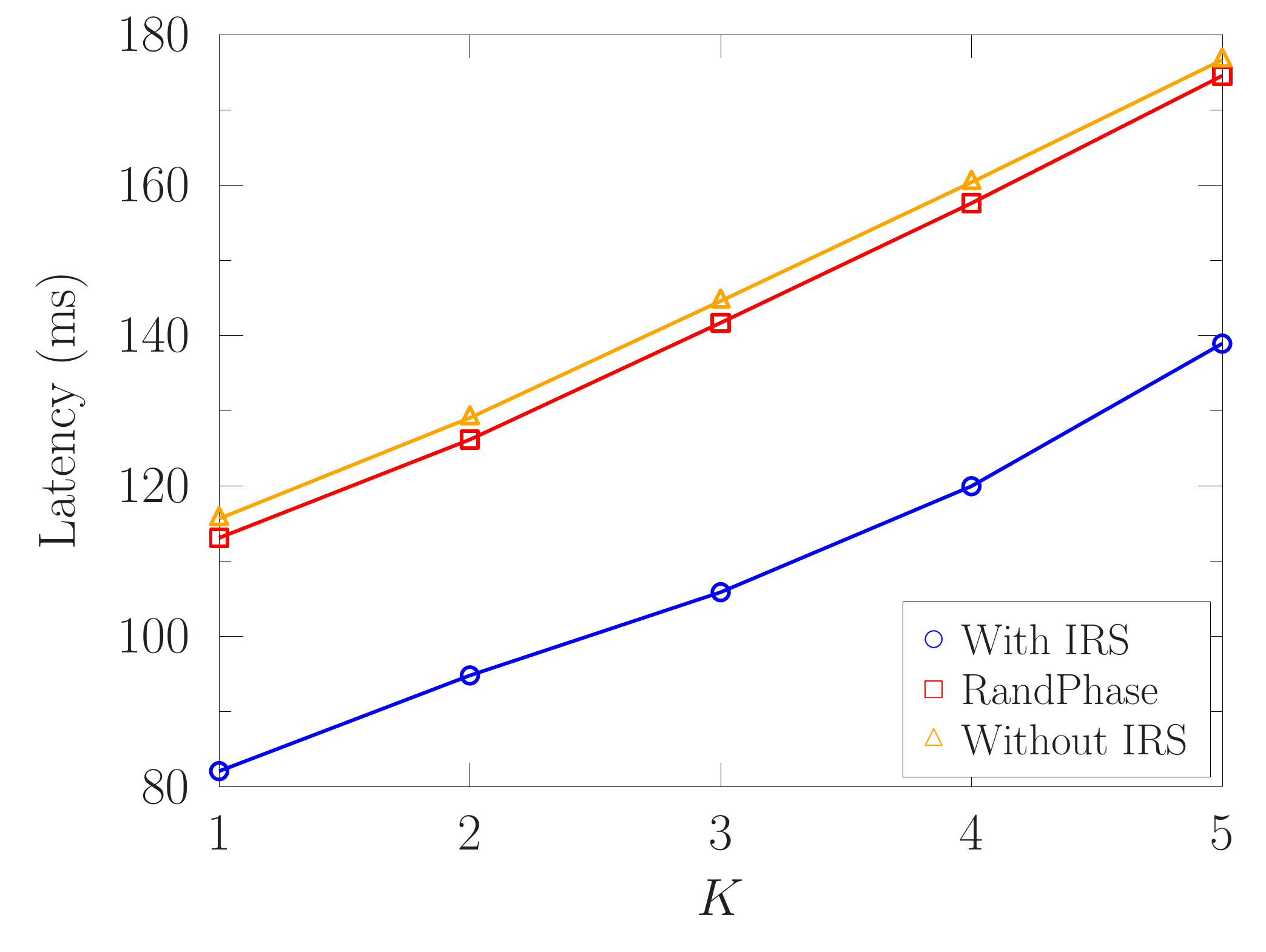}
  \captionof{figure}{Simulation results of the device-average latency versus the number of devices $K$. The parameters are set as follows: $d = 280~\rm{m}$, $\overline{d} = 10~\rm{m}$, $r = 10~\rm{m}$, $N = 40$, and $f^e_{\text{total}} = 50\times 10^9~\rm{cycle/s}$.}
  \label{fig:mu_mu}
\end{figure}

\subsubsection{Impact of the Number of Devices}
Fig.~\ref{fig:mu_mu} shows the latency versus the number of devices in the cycle in multi-device scenario. It can be readily observed that the device-average latency increases upon increasing the number of devices in the IRS-aided MEC system. This is partially because of the reduced edge computational resources allocated to each device and partially due to the reduced beamforming gain achieved at each device. The former issue may be overcome by equipping the edge node with more powerful computing capability, while the latter problem can be solved by deploying more IRSs in the MEC system for forming stronger beams. Nonetheless, compared to the ``Without IRS" scheme, our ``With IRS" scheme is capable of reducing the device-average latency from 177 ms to 139 ms, when we have $5$ devices in the MEC system. This again validates the benefits of our proposed system.

\begin{figure}[t!]\center
  \includegraphics[width=0.47\textwidth]{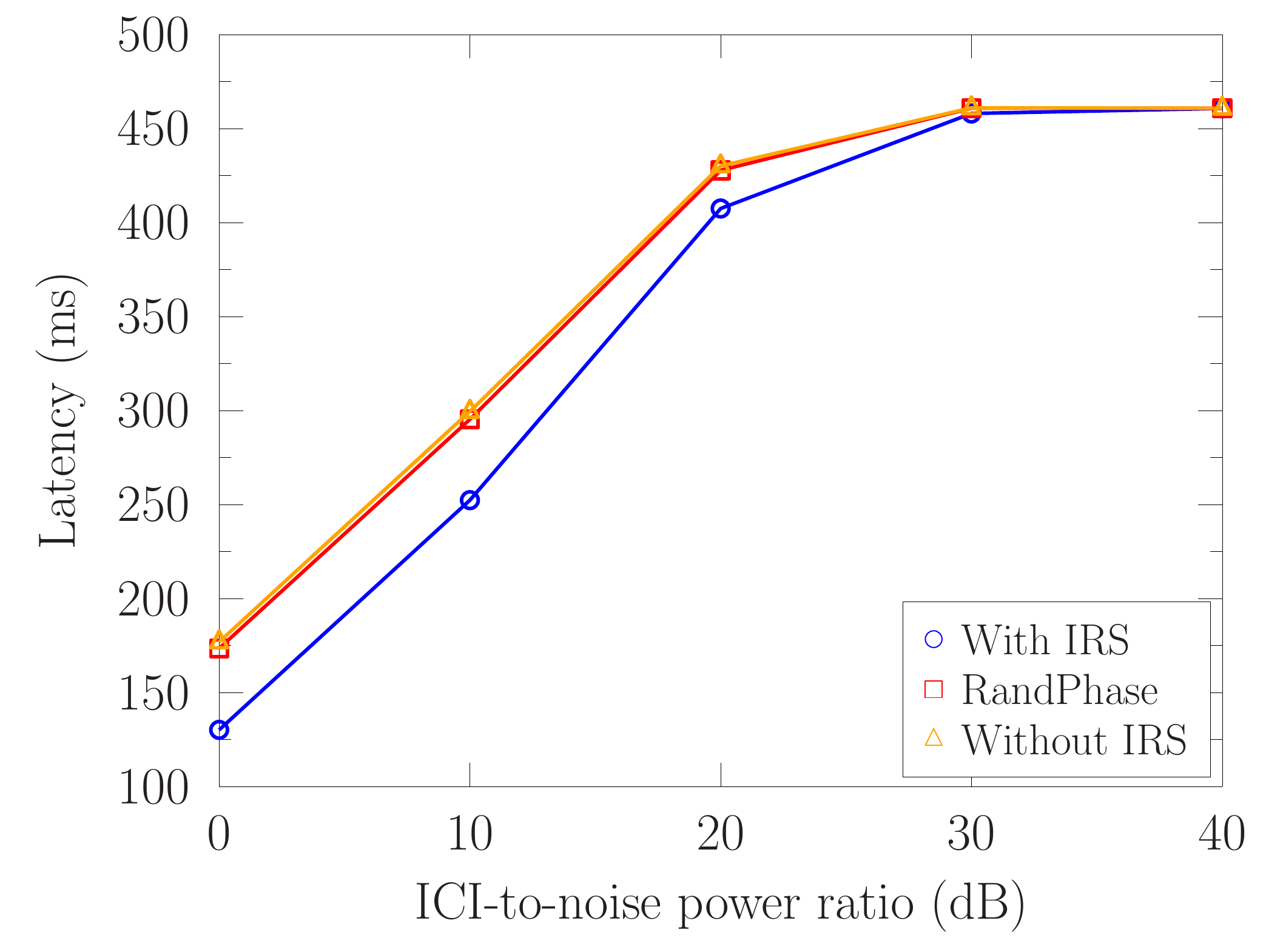}
  \captionof{figure}{Simulation results of the device-average latency versus the ICI-to-noise ratio. The parameters are set as follows: $K = 3$, $d = 280~\rm{m}$, $\overline{d} = 10~\rm{m}$, $r = 10~\rm{m}$, $N = 40$, and $f^e_{\text{total}} = 50\times 10^9~\rm{cycle/s}$.}
  \label{fig:mu_ICI}
\end{figure}
\subsubsection{Impact of Inter-Cell Interference}
In realistic scenarios, inter-cell interference (ICI) also degrades computation off-loading. To quantify the impact of ICI, Fig.~\ref{fig:mu_ICI} presents the latency versus the ICI-to-noise power ratio, where the BS is assumed to know the power of the received interference but not the specific signal transmitted from other cells. Observe that the benefit of employing IRSs in MEC systems decreases upon increasing the ICI-to-noise power ratio. To elaborate, when computation off-loading is used in the face of strong ICI, the fraction of tasks that can be off-loaded becomes marginal. In this case, the wireless devices have to rely on their own computing capabilities. In other words, the potential of IRSs may not be fully exploited. This observation suggests an important insight for engineering design: the spectrum allocation of adjacent cells has to be carefully managed for minimizing the ICI in IRS-aided MEC systems.

\section{Conclusions}
\label{sec:Conclusions}

In order to reduce the computational latency, an IRS was proposed for employment in MEC systems. Based on this model, a latency-minimization problem was formulated, subject to practical constraints on the total edge computing capability and IRS phase shifts. Sophisticated algorithms were developed for optimizing both the computing and communications settings.
The benefits of using IRSs in the MEC system were evaluated under various simulation environments. Quantitatively, the device-average computational latency was reduced from $177~\rm{ms}$ to $139~\rm{ms}$, compared to the conventional MEC system operating without IRSs in a single cell associated with the cell radius of $300~\rm{m}$, a $5$-antenna access point and $5$ active devices.
Furthermore, the rapid convergence of our proposed algorithm was confirmed numerically, which validates their benefits.
As our future work, an energy-minimization based design will be conceived for IRS-aided MEC systems.

\ifCLASSOPTIONcaptionsoff
  \newpage
\fi

\appendices

\section{The proof of Proposition~\ref{proposition:offloading_volume}}
\label{app:Proposition_1}

$\hat{\ell}_k \in [0,L_k]$ is used to represent the relaxation \cite{fisher1981lagrangian} of the integer value $\ell_k \in \{0,1,\ldots , L_k\}$. Furthermore, given the values of $\pmb{W}, \pmb{\theta}$ and $\pmb{f}^e$, we define the delay associated with $\hat{\ell}_k$ to be $\hat{D}_k(\hat{\ell}_k) \triangleq \max \big\{ D^l_k(\hat{\ell}_k), D^e_k(\hat{\ell}_k) \big\}$, which can be reformulated from \eqref{eqn:latency} as a segmented form below
\begin{equation}
\label{eqn:latency_piecewise}
\hat{D}_k(\hat{\ell}_k) = \begin{cases}
    \frac{(L_k-\hat{\ell}_k) c_k}{f^l_k}, & 0 \leq \hat{\ell}_k \leq \frac{L_k c_k R_k f^e_k}{f^e_kf^l_k + c_k R_k (f^l_k+f^e_k)}, \\
    \frac{\hat{\ell}_k}{R_k} + \frac{\hat{\ell}_k c_k}{f^e_k}, & \frac{L_k c_k R_k f^e_k}{f^e_kf^l_k + c_k R_k (f^l_k+f^e_k)} < \hat{\ell}_k \leq L_k.
  \end{cases}
\end{equation}
A glance at \eqref{eqn:latency_piecewise} reveals that $\hat{D}_k(\hat{\ell}_k)$ decreases upon increasing $\hat{\ell}_k$ in the range of $\hat{\ell}_k \in \bigg[0,\frac{L_k c_k R_k f^e_k}{f^e_k f^l_k + c_kR_k \big(f^e_k + f^l_k \big)} \bigg]$, while $\hat{D}_k(\hat{\ell}_k)$ increases upon increasing $\hat{\ell}_k$ in the range of $\hat{\ell}_k \in \bigg[\frac{L_k c_k R_k f^e_k}{f^e_k f^l_k + c_kR_k \big(f^e_k + f^l_k \big)}, L_k \bigg]$. Therefore, it is readily inferred that $\hat{D}_k(\hat{\ell}_k)$ achieves its minimum value when we set $\hat{\ell}_k = \frac{L_k c_k R_k f^e_k}{f^e_k f^l_k + c_kR_k \big(f^e_k + f^l_k \big)}$, which is denoted by $\hat{\ell}_k^*$. Bearing in mind that the optimal value of $\ell_k$ has to be an integer, it may be obtained by carrying out the operation $\ell^*_k = \mathop{\arg \min} \limits_{\hat{\ell} \in \big\{ \lfloor \hat{\ell}^*_k \rfloor, \lceil \hat{\ell}^*_k\rceil \big\} } D_k (\hat{\ell}_k)$. This completes the proof.

\section{The proof of Proposition~\ref{proposition:convex_optimization}}
\label{app:Proposition_2}
Let us denote the second derivative of the OF of Problem $\mathcal{P}1\text{-}E$ with respect to $f^e_k$ by $\Phi_{1\text{-}E}$, which is calculated as
\begin{IEEEeqnarray}{rCl}
\Phi_{1\text{-}E} = \frac{2\varpi_k L_k c_k^3 R_k^2 (f^l_k + c_k R_k)}{\big[f^e_k f^l_k + c_kR_k \big(f^e_k + f^l_k \big) \big]^3}.
\end{IEEEeqnarray}
Since the values of $\varpi_k$, $c_k$, $R_k$, $f^l_k$ are all positive and we have $L_k \geq 0$, $f^e_k \geq 0$,  it may be readily demonstrated that $\Phi_{1\text{-}E} \geq 0$. Hence the OF is a convex function with respect to $f^e_k$. Furthermore, the constraint functions \eqref{eqn:P0_constraint3} and \eqref{eqn:P0_constraint4} are all of linear forms. Hence, Problem $\mathcal{P}1\text{-}E$ is shown to be a strictly convex problem.

\section{The proof of Proposition~\ref{Proposition:problem_transfer}}
\label{app:Proposition_3}
The Lagrangian of Problem $\mathcal{P}2\text{-}E2$ is given by
\begin{IEEEeqnarray}{rCl}
\mathcal{L}(\pmb{W},\pmb{\theta}, \pmb{\beta}, \pmb{\lambda}) = \sum^K_{k=1} \beta_k + \sum^{K}_{k = 1} \lambda_k \big[ \varpi_k \ell_k -\beta_k R_k(\pmb{w}_k,\pmb{\theta}) \big]
\end{IEEEeqnarray}
where $\{ \lambda_k \}$ is the non-negative Lagrange multiplier.
If $(\pmb{W}^*,\pmb{\theta}^*,\pmb{\beta}^*)$ is the solution of Problem $\mathcal{P}2\text{-}E2$, there exists $\pmb{\lambda}^*$ satisfying the following KKT conditions
\begin{align}
& \frac{\partial \mathcal{L}}{\partial \theta_k} = - \lambda_k^* \beta_k^* \triangledown R_k (\pmb{w}_k^*,\pmb{\theta}^*) = 0, \quad  k = 1,2,\ldots,K, \label{eqn:KKT_2_E_1} \\
& \frac{\partial \mathcal{L}}{\partial \pmb{w}_k} = - \lambda_k^* \beta_k^* \triangledown R_k (\pmb{w}_k^*,\pmb{\theta}^*) = 0, \quad k = 1,2,\ldots,K, \label{eqn:KKT_2_E_2} \\
& \frac{\partial \mathcal{L}}{\partial \beta_k} = 1 - \lambda_k^* R_k(\pmb{w}_k^*,\pmb{\theta}^*) = 0, \quad k = 1,2,\ldots,K, \label{eqn:KKT_2_E_3} \\
& \lambda_k^* \big[ \varpi_k \ell_k -\beta_k^* R_k(\pmb{w}_k^*,\pmb{\theta}^*) \big] = 0, \quad k = 1,2,\ldots,K, \label{eqn:KKT_2_E_4} \\
& \lambda_k^* \geq 0, \quad k = 1,2,\ldots,K, \label{eqn:KKT_2_E_5} \\
& \varpi_k \ell_k -\beta_k^* R_k(\pmb{w}_k^*,\pmb{\theta}^*) \leq 0, \quad k = 1,2,\ldots,K, \label{eqn:KKT_2_E_6} \\
& 0 \leq \theta_k^* \leq 2\pi,  \quad k = 1,2,\ldots,K. \label{eqn:KKT_2_E_7}
\end{align}
Since we have $R_k(\pmb{w}_k,\pmb{\theta}) > 0$, \eqref{eqn:KKT_2_E_3} is equivalent to
\begin{IEEEeqnarray}{rCl}
\lambda_k^* = \frac{1}{R_k(\pmb{w}_k^*,\pmb{\theta}^*)}, \quad \forall k \in \{1,2,\ldots,K\},
\end{IEEEeqnarray}
and then \eqref{eqn:KKT_2_E_4} is equivalently written as
\begin{IEEEeqnarray}{rCl}
\beta_k^* = \frac{\varpi_k \ell_k}{R_k(\pmb{w}_k^*,\pmb{\theta}^*)}, \quad \forall k \in \{1,2,\ldots,K\}.
\end{IEEEeqnarray}
Furthermore, Eq. \eqref{eqn:KKT_2_E_1}, \eqref{eqn:KKT_2_E_2} and \eqref{eqn:KKT_2_E_7} are exactly the KKT conditions of Problem $\mathcal{P}2\text{-}E3$, when we set $\pmb{\lambda} = \pmb{\lambda}^*$ and $\pmb{\beta} = \pmb{\beta}^*$. This proves the first conclusion of Proposition \ref{Proposition:problem_transfer}. Following the same procedure, the second conclusion of Proposition \ref{Proposition:problem_transfer} may also be readily shown.

\bibliographystyle{ieeetr}
\bibliography{IEEEabrv.bib}

\end{document}